\documentclass[aps,twocolumn,showpacs,amsmath,amssymb,floatfix,eqsecnum]{revtex4}
\usepackage{graphicx}
\usepackage{dcolumn}
\usepackage{bm}

\begin{document}

\title{The critical Casimir force and its fluctuations in \\
lattice spin models: exact and Monte Carlo results}
\author{Daniel Dantchev$^{1,2,3}$\thanks{e-mail:
danield@mf.mpg.de} and Michael Krech$^{2,3}$\thanks{e-mail:
mkrech@mf.mpg.de}
}
\affiliation{
$^1$Institute of Mechanics - BAS, Acad. G. Bonchev St. bl. 4,
1113 Sofia, Bulgaria,\\
$^2$Max-Planck-Institut f\"{u}r Metallforschung,
Heisenbergstrasse 1, D-70569 Stuttgart, Germany,\\
$^3$Institut f\"{u}r Theoretische und Angewandte Physik,
Universit\"{a}t Stuttgart, Pfaffenwaldring 57, D-70569 Stuttgart,
Germany}

\begin{abstract}
We present general arguments and construct a stress tensor
operator for finite lattice spin models. The average value of this
operator gives the Casimir force of the system close to the bulk
critical temperature $T_c$. We verify our arguments via exact
results for the force in the two-dimensional Ising model,
$d$-dimensional Gaussian and mean spherical model with $2<d<4$. On
the basis of these exact results and by Monte Carlo simulations
for three-dimensional Ising, XY and Heisenberg models we
demonstrate that the standard deviation of the Casimir force $F_C$
in a slab geometry confining a critical substance in-between is
$k_b T D(T)(A/a^{d-1})^{1/2}$, where $A$ is the surface area of
the plates, $a$ is the lattice spacing and $D(T)$ is a slowly
varying nonuniversal function of the temperature $T$. The
numerical calculations demonstrate that at the critical
temperature $T_c$ the force possesses a Gaussian distribution
centered at the mean value of the force $<F_C>=k_b T_c
(d-1)\Delta/(L/a)^{d}$, where $L$ is the distance between the
plates and $\Delta$ is the (universal) Casimir amplitude.

\end{abstract}
\pacs{64.60.-i, 64.60.Fr, 75.40.-s}

\maketitle

\section{Introduction}

If material bodies are immersed in a fluctuating medium the surfaces
of these bodies impose boundary conditions that select a certain mode
spectrum for the fluctuations. This leads to a contribution into the
ground state energy of a quantum mechanical system, or to the
free energy of a critical statistical mechanical system, which
depends on the geometrical parameters characterizing the mutual
position of the bodies and their shape. This is known as the
Casimir effect \cite{C48,C53,FG78}.

According to our present understanding, the Casimir effect is a
phenomenon common to all systems characterized by fluctuating
quantities on which external boundary conditions are imposed.
Casimir forces arise from an interaction between distant portions
of the system mediated by fluctuations.

In quantum mechanics one usually considers fluctuations of the
electromagnetic field. In this case correlations of the fluctuations
are mediated by photons -- massless excitations of the
electromagnetic field \cite{C48,C53,MT97,M01}. In statistical
mechanics the massless excitations can be generated by critical
fluctuations of the order parameter around the critical
temperature $T_c$ of the system \cite{FG78,K94,BDT00}. Goldstone modes
(or spin wave excitations) in $O(n)$ models at temperatures below $T_c$
also provide massless excitations \cite{APP91,ZPZ00,KG99}. Fluctuations
of this type are scale invariant and therefore the Casimir force is long
ranged in the above cases.

In this article we discuss the behavior of the thermodynamic
Casimir force in systems with short-ranged interactions undergoing
a second order phase transition.

To be more specific, let us consider a statistical mechanical
system, a magnet or a fluid, with the slab geometry
$L_\parallel^{d-1}\times L_\perp$, where $d$ is the dimensionality
of the system and periodic boundary conditions are applied. In
the limit $L_\parallel \rightarrow \infty$ ($L_\perp$ fixed) the
Casimir force per unit area is defined as
\begin{equation}
\beta F_{\rm Casimir} (T,L_\perp)=-\frac{\partial f_{{\rm
ex}}(T,L_\perp)}{\partial L_\perp}, \label{def}
\end{equation}
where $f _{{\rm ex}}(T,L_\perp)$ is the excess free energy
\begin{equation}
f _{{\rm ex}}(T,L_\perp)=f (T,L_\perp)-L_\perp f_{{\rm bulk}}(T)
\label{fexd}
\end{equation}
of the system.
Here $f(T,L_\perp)$ is the full free energy per unit area measured in units
of $k_BT$ and $f_{{\rm bulk}}(T)$ is the corresponding bulk free energy
density.

According to the definition given by Eq.(\ref{def}) the
thermodynamic Casimir force is a generalized force conjugate to
the distance $L_\perp$ between the boundaries of the system with
the property $F_{{\rm Casimir}}(T,L_\perp)\rightarrow 0$ for
$L_\perp\rightarrow \infty$. We are interested in the behavior of
$F_{\rm Casimir}$ when $L_\perp\gg a$, where $a$ is a typical
microscopic length scale. In this limit finite size scaling theory
is applicable. Then one has \cite{privman90}
\begin{equation}
\beta F_{{\rm Casimir}}(T,L_\perp)=L_\perp^{-d}X_{\rm
Casimir}(L/\xi_\infty),
\end{equation}
where $\xi_\infty$ is the true bulk correlation length, while
$X_{\rm Casimir}$ is an universal scaling function.

At the critical point $T_c$ of the bulk system one has
$\xi_\infty=\infty$, and \cite{FG78}
\begin{equation}
\label{cadef} \beta_c F_{\rm Casimir}
(T_c,L_\perp)=(d-1)\frac{\triangle}{L_\perp^d},
\end{equation}
where $\triangle$ is the so-called Casimir amplitude. This
amplitude  is universal, i.e. $\triangle$ depends only on the
universality class of the corresponding bulk system and the type
of boundary conditions used across $L_\perp$. Obviously, one has
$\triangle=X_{\rm Casimir}(0)/(d-1)$.

The Casimir force may also be viewed from the point of view of
conformal invariance of, e.g., critical systems \cite{Ca87}. The
Casimir force in its simple form for the film geometry
($L_\parallel \to \infty$, $L_\perp$ finite) is due to the
$L_\perp$ dependence of the free energy $f(T,L_\perp)$ per unit
area. The free energy therefore responds to any coordinate
transformation which changes the value of $L_\perp$. On the other
hand any coordinate transformation which transforms a slab of
thickness $L_\perp$ into a slab of a different thickness is
nonconformal. Therefore, the Casimir force is the response of the
free energy $f(T,L_\perp)$ of the original slab to a nonconformal
coordinate transformation. The change of the free energy due to a
nonconformal coordinate transformation is determined by the
thermal average of the stress tensor $t_{\alpha \beta}$ associated
with the Hamiltonian of the system \cite{Ca87}. If the coordinate
perpendicular to the surfaces of the slab is denoted by $z$, it is
easy to prove, that the Casimir force in the slab is given by the
thermal average of the stress tensor component $t_{zz}$
\cite{Ca87} (see below).

Normally, one is interested in $<t_{zz}>$, which determines the
(average) value of the Casimir force. In addition, one can
consider any realization $t_{zz}$ of the stress tensor to be
proportional to any realization, e.g. instantaneous value of the
Casimir force $F_C$, where  $ <~F_C~> \equiv F_{Casimir}$. That is
the approach undertaken recently by Bartolo, Ajdari, Fournier and
Golestenian \cite{BAFG02}. They consider a statistical-mechanical
model of a $d$-dimensional medium described by a scalar field
$\Phi$ with an elastic energy density proportional to
$(\overrightarrow{\nabla} \Phi)^2$, i.e. one considers an elastic
Hamiltonian of the form
\begin{equation}
{\cal H}[\Phi]=\frac{K}{2}\int d^d {\bf
R}\left[\left(\overrightarrow{\nabla} \Phi\left({\bf
R}\right)\right)^2\right],
\end{equation}
where ${\bf R}=({\bf r},z)$. They assume that the plates impose
Dirichlet boundary conditions, i.e. $\Phi({\bf r},0)=\Phi({\bf
r},L_\perp)=0$.

 For the total Casimir force $F_C^t\equiv
L_\parallel ^{d-1} F_C$ it has been found that:
\begin{equation}
\label{mean} \langle F_C^t \rangle \equiv F^t_{\rm Casimir}=c(d)
\frac{L_\parallel^{d-1}}{\beta L_\perp^d}=c(d)\frac{A}{\beta
L_\perp^d},
\end{equation}
where $c(d)=(d-1)\Gamma(d/2)\zeta(d)/(4\pi)^{d/2}$ depends only on
the spacial dimension $d$, while the variance of the force
$(\triangle F_C^t)^2$, that can be considered of being produced of
$N_a=(L_\parallel/a)^{d-1}$ independent strings is
\begin{equation}
\label{var}
 (\triangle F_C^t)^2 \propto \frac{1}{\beta^2}
\left(\frac{L_\parallel}{a}\right)^{d-1} = \frac{1}{\beta^2}
\frac{A}{a^{d-1}}.
\end{equation}
In the above expressions $A \equiv L_\parallel^{d-1}$ is the
cross-section of the system. From Eqs. (\ref{mean}) and
(\ref{var}) one obtains that the "noise-over-signal ratio" $\rho$
is
\begin{equation}
\rho \equiv \frac{\triangle F_C^t}{<F_C^t>}\propto
\left(\frac{L_\perp}{L_\parallel}\right)^{(d-1)/2}\left(\frac{L_\perp}{a}\right)^{(d+1)/2}.
\end{equation}
The probability distribution of the force has been found to be
Gaussian, i.e.
\begin{equation}
{\cal P}(F_C^t=x)=\frac{1}{\sqrt{2\pi \triangle
F_C^t}}\exp\left[-\frac{(x-<F_C^t>)^2}{2(\triangle F_C^t)^2}
\right].
\end{equation}

The structure of the current article is as follows. First,  in Section \ref{Secstdef} we present some general arguments and construct a stress tensor operator for finite {\it lattice} spin models. Then, in Section  \ref{AR}, we verify our arguments by presenting exact results for the two-dimensional Ising model (see Section \ref{S2dIsing}), $d$-dimensional Gaussian model  (see Section \ref{GM}), and for the mean spherical model with $2<d<4$ (see Section \ref{SM}). Monte Carlo results, based on our definition, are given in Section \ref{SMCR} for the behavior of the force and its variance. There the three-dimensional Ising (Section \ref{S3dIsing}), XY (Section \ref{S3dXYM}) and Heisenberg (Section \ref{S3dHM}) models have been considered. The article closes with a Discussion given in Section \ref{SSu}. The set of technical details needed in the main text is organized in a series of Appendixes.


\section{The stress tensor for lattice spin models}
\label{Secstdef}

We will now reconsider the Casimir force for a $d$-dimensional
anisotropic lattice $O(N)$ spin system with the Hamiltonian (see
also Ref.\cite{BK98})
\begin{equation} \label{Hamlam}
{\cal H}(\lambda)=-\sum_{\bf R} \sum_{k=1}^d J_k(\lambda) S_{\bf
R} S_{{\bf R}+{\bf e}_k}
\end{equation}
where a $d$-dimensional simple hypercubic lattice with
$L_\parallel^{d-1} \times L_\perp$ lattice sites and $L_\parallel
\gg L_\perp$ is assumed. The vector $\bf R$ indicates a lattice
site and the vectors ${\bf e}_k$, $k = 1, 2, \dots, d$ connect
nearest neighbor lattice sites on the simple hypercubic lattice.
The spins $S_{\bf R}$ are considered to be of $O(N)$ type.
Following Ref.\cite{BK98} and the general idea of conformal field
theory \cite{Ca87}, we define the coupling constants in
Eq.(\ref{Hamlam}) by
\begin{eqnarray} \label{Jlam} J_k(\lambda) &\equiv&
J_\parallel(\lambda) = J\left(e^\lambda\right),\ k = 1, \dots,
d-1,
\nonumber \\
J_d(\lambda) &\equiv& J_\perp(\lambda) =
J\left(e^{-(d-1)\lambda}\right),
\end{eqnarray}
which means that $\lambda = 0$ marks the isotropic point of
Eq.(\ref{Hamlam}) due to $J_\parallel(0) = J_\perp(0) = J(1)$. The
function $J(x)$ in Eq.(\ref{Jlam}) is supposed to be smooth and
monotonic in the vicinity of $x = 1$ but is otherwise arbitrary.
The critical point of the bulk spin model defined by
Eq.(\ref{Hamlam}) is given by an implicit equation of the type
\begin{eqnarray} \label{Crilam}
{\cal K}(\beta_c J_1(\lambda), \dots, \beta_c J_{d-1}(\lambda),
\beta_c J_d(\lambda)) = \nonumber \\
{\cal K}(\beta_c J_\parallel(\lambda), \dots, \beta_c
J_\parallel(\lambda), \beta_c J_\perp(\lambda)) = 1,
\end{eqnarray}
where $\beta_c = 1 /(k_B T_c)$. The function ${\cal K}(u_1, \dots,
u_{d-1}, u_d)$ is a smooth function of $d$ variables $u_1, \dots,
u_d$. Furthermore, at ${\cal K}(u_1, \dots, u_d) = 1$ the function
${\cal K}$ is invariant with respect to any permutation of its
arguments, because the location of the critical point is
independent of the labelling of the lattice axes. This also
implies, that the derivatives $\partial {\cal K} / \partial u_k$,
$k = 1, \dots, d$ for ${\cal K} = 1$ all have the same value
${\cal K}'\neq 0$ at the isotropic point $u_1 = u_2 = \dots =
u_d$. For the two dimensional Ising model the function $\cal K$ is
rigorously known \cite{CW73}
\begin{equation} \label{Cri2dI}
{\cal K}(u_1, u_2) = \sinh(2 u_1) \sinh(2 u_2),
\end{equation}
but for $d \geq 3$ exact results for ${\cal K}$ are extremely
rare. From Eq.(\ref{Crilam}) one immediately concludes, that in
general the critical temperature $T_c$ will depend on the
anisotropy parameter $\lambda$. However, for the particular
parameterization given by Eq.(\ref{Jlam}) one finds at the
critical point (see also Ref.\cite{BK98})
\begin{eqnarray}\label{Tc0}
0 = \left.\frac{d {\cal K}}{d\lambda}\right|_{\lambda=0} &=&
\beta_c'(0) {\cal K}'\left[ (d-1) J_\parallel(0) + J_\perp(0)
\right]
\nonumber \\
&+& \beta_c(0) {\cal K}'\left[ (d-1) J_\parallel'(0) + J_\perp'(0) \right] \\
&=& \beta_c'(0) {\cal K}' d J(1), \nonumber
\end{eqnarray}
which immediately yields $\beta_c'(0) = 0$, i.e., for an
infinitesimal anisotropy $(\lambda \ll 1)$ the value of the
critical temperature remains unchanged and is given by its value
for the isotropic model.

The correlation length of the anisotropic bulk spin system defined
by Eqs.(\ref{Hamlam}) and (\ref{Jlam}) will also be anisotropic.
In the vicinity of the (isotropic) bulk critical temperature $T_c$
one finds
\begin{equation} \label{xilam}
\xi_\parallel(\lambda,t) = \xi_{\parallel,0}(\lambda) |t|^{-\nu}
\quad \mbox{and} \quad \xi_\perp(\lambda,t) =
\xi_{\perp,0}(\lambda) |t|^{-\nu},
\end{equation}
where $t = (T - T_c)/T_c$ is the reduced temperature and $\nu$ is
the correlation length exponent which is universal, i.e.,
independent of the anisotropy parameter $\lambda$. At the
isotropic point $\lambda = 0$ one has $\xi_{\parallel,0}(0) =
\xi_{\perp,0}(0) \equiv \xi_0$. To simplify the notation we ignore
the fact that the correlation length amplitudes in general also
depend on the sign of the reduced temperature $t$. We therefore
assume that $t > 0$ in the following.

In order to be able to apply finite-size scaling in the critical
regime with respect to a {\em single} correlation length, say,
$\xi_\parallel$, we employ the following anisotropic rescaling of
the spatial coordinates:
\begin{equation} \label{coorlam}
x'_k = x_k, \ k = 1, \dots, d-1, \quad \mbox{and} \quad x'_d =
\frac{\xi_{\parallel,0}(\lambda)}{\xi_{\perp,0}(\lambda)}\ x_d.
\end{equation}
This transformation has the desired property, namely
$\xi_\parallel' = \xi_\parallel$ and $\xi_\perp' = \xi_\parallel$
as can be easily verified from Eqs.(\ref{xilam}) and
(\ref{coorlam}). For the lattice sizes $L_\parallel$ and $L_\perp$
we find accordingly
\begin{equation} \label{Llam}
L_\parallel' = L_\parallel \quad \mbox{and} \quad L_\perp' =
\frac{\xi_{\parallel,0}(\lambda)}{\xi_{\perp,0}(\lambda)}\
L_\perp \equiv R(\lambda)\ L_\perp.
\end{equation}
The link between the explicit $\lambda$-dependence of the free
energy of our spin system according to Eqs.(\ref{Hamlam}) and
(\ref{Jlam}) and the Casimir force defined by Eq.(\ref{def}) is
provided by Eq.(\ref{Llam}). In the limit $L_\parallel \to \infty$
we find (see also Eqs.(24) and (33) of Ref. \cite{BK98})
\begin{eqnarray} \label{fexlam}
\left.\frac{d f_{\rm ex}}{d\lambda} \right|_{\lambda=0} &=&
-\lim_{L_\parallel \to \infty} \frac{\beta J'(1)}{L_\parallel^{d-1}}\\
&\times& \left\langle \sum_{\bf R} \left[\sum_{k=1}^{d-1} S_{\bf
R} S_{{\bf R}+{\bf e}_k} - (d-1) S_{\bf R} S_{{\bf R}+{\bf e}_d}
\right]\right\rangle, \nonumber
\end{eqnarray}
where $\langle \dots \rangle$ denotes the thermal average with
respect to the Hamiltonian given by Eq.(\ref{Hamlam}) at the
isotropic point $\lambda=0$. From Eqs.(\ref {def}) and (\ref{fexlam})
and the finite size scaling form
\begin{equation}\label{fexsingscal}
f_{\rm ex}(t,L_\perp) = L_\perp^{-(d-1)}g_{\rm ex}\left[\ t\
\left(L_\perp / \xi_{\perp,0}\right)^{1/\nu} \right]
\end{equation}
of the excess free energy $f_{ex}(t,L_\perp)$ in the limit
$L_\parallel \to \infty$, we obtain an expression of the Casimir
force which is derived in detail in Appendices \ref{clar} and
\ref{ast}. From Eq.(\ref{FCasimir}) derived in Appendix \ref{ast}
the operator form of the stress tensor component $t_{\perp
\perp}({\bf R})$ can be read off as
\begin{eqnarray}
\label{st} t_{\perp \perp}({\bf R})&=&\beta
J'(1)\left[R'(0)\right]^{-1} \nonumber\\
& & \times \left[\sum_{k=1}^{d-1} S_{\bf R} S_{{\bf R}+{\bf e}_k}
- (d-1) S_{\bf R} S_{{\bf R}+{\bf e}_d}\right] \nonumber \\
&&  +\frac{1}{d\nu}({\cal \hat{H}}-{\cal
\hat{H}}_b)(\beta_c(0)-\beta),
\end{eqnarray}
where Eq.(\ref{uub}) was used and the operators ${\cal \hat{H}}$
and ${\cal \hat{H}}_b$ are properly normalized Hamiltonians ${\cal
H}(0)$ (see Eq.(\ref{Hamlam}) and Appendix \ref{ast}).

Eq.(\ref{st}) provides the connection between
the stress tensor component $t_{\perp \perp}$ parallel
to the surfaces of the slab and the spin lattice model given by
Eq.(\ref{Hamlam}). It is valid also for temperatures $T > T_c$ and
thus generalizes Eq.(36) of Ref.\cite{BK98}. For $T < T_c$
Eq.(\ref{st}) holds also for $O(N)$ symmetric spin models,
because the correlation length ratio $\xi_\parallel / \xi_\perp$
remains finite at $T = T_c$ and it can be continued analytically
into the Goldstone regime, where it can be used for the
anisotropic rescaling of the coordinates according to
Eq.(\ref{coorlam}). Finally, we note that Eq.(\ref{st}) only
holds for periodic boundary conditions.

For the purposes of this investigation Eq.(\ref{st}) serves
as a prescription to obtain the universal scaling function of the
Casimir force in critical slabs with periodic boundary conditions
by Monte-Carlo simulations. However, Eq.(\ref{st}) contains
the ratio $\xi_{\parallel,0}(\lambda) / \xi_{\perp,0}(\lambda)$ of
the correlation length amplitudes for Eq.(\ref{Hamlam}) as a
prefactor. For the two dimensional Ising model this function is
given by \cite{CW73}
\begin{equation} \label{xip0p0}
\frac{\xi_{\parallel,0}(\lambda)}{\xi_{\perp,0}(\lambda)} =
\frac{J_\perp(\lambda)+J_\parallel (\lambda) \sinh
[2\beta_c(\lambda) J_\perp(\lambda)]}{J_\parallel
(\lambda)+J_\perp (\lambda) \sinh [2\beta_c(\lambda)
J_\parallel(\lambda)]}
\end{equation}
and by virtue of Eq.(\ref{Jlam}) Eq.(\ref{st}) can be made
explicit. But in $d \geq 3$ no such information is available.
However, in the critical regime, where Eq.(\ref{st}) will be
applied, the scaling argument $L_\perp / \xi$ will typically not
exceed values of the order 10. This means that we will be dealing
with reduced temperatures in the range $|t| < 10 (L_\perp /
\xi_0)^{-1/\nu}$, i.e., the relevant temperature range diminishes
as the system size $L_\perp$ increases. As can be seen explicitly
in Eq.(\ref{xip0p0}) the correlation length amplitude ratio only
depends on the temperature $T$ and does {\em not} display any
scaling behavior. Therefore, the generally unknown prefactor in
Ep.(\ref{st}) can be treated as a constant for sufficiently
large system sizes which can be determined by a normalization of
$\langle t_{\perp \perp} \rangle$ to known results at $T = T_c$
\cite{K94}.

We end this section with some observations that turn out to be
very helpful for analytical calculations of the variance of the
force. Since they are model independent we give them before
passing to explicit calculations presented in the next section.
Lets us consider an anisotropic Hamiltonian of the type given by
Eq. (\ref{Hamlam}), where $J_1(\lambda)=\cdots =
J_{d-1}(\lambda)=J_{\parallel}(\lambda)=(1+\lambda)J$ and
$J_d(\lambda)=J_{\perp}(\lambda)=(1-(d-1)\lambda)J$. Then, it is
easy to see that
\begin{equation}\label{htt}
   {\cal H}(\lambda)={\cal H}(0)-\lambda J \sum_{\bf R}\left[
\sum_{k=1}^{d-1} S_{\bf R}S_{{\bf R}+{\bf e}_k} -(d-1)S_{\bf
R}S_{{\bf R}+{\bf e}_d} \right],
\end{equation}
or, equivalently,
\begin{equation}\label{htdef}
   \beta {\cal H}(\lambda) = \beta {\cal H}(0)-\lambda \sum_{\bf
R}\tilde{t}_{\perp,\perp}({\bf R}),
\end{equation}
where
\begin{equation}\label{tt}
\tilde{t}_{\perp,\perp}({\bf R})=\beta J \left[ \sum_{k=1}^{d-1}
S_{\bf R}S_{{\bf R}+{\bf e}_k}-(d-1)S_{\bf R}S_{{\bf R}+{\bf
e}_d}\right]
\end{equation}
differs only by a multiplying factor from the stress tensor
$t_{\perp,\perp}({\bf R})$ defined in Eq. (\ref{st}).

Let $\tilde{f}(T,\lambda)$ is the total free energy per unit spin
of a system with the Hamiltonian (\ref{htdef}). Then, taking into
account the translational invariance symmetry, it is easy to see
that
\begin{eqnarray}\label{mvt}
\langle \tilde{t}_{\perp,\perp}({\bf R})\rangle &=& \left.
-\frac{\partial}{\partial \lambda}\left[ \beta
\tilde{f}(T,\lambda)\right]\right|_{\lambda=0}  \\ \nonumber &=&
\beta J (d-1) \left[\langle S_{\bf 0}S_{{\bf e}_1}\rangle-\langle
S_{\bf 0}S_{{\bf e}_d} \rangle \right].
\end{eqnarray}
Therefore, in order to calculate the average value of
$\tilde{t}_{\perp,\perp}({\bf R})$ one needs either to know the
finite-size free energy density of an anisotropic system, or, what
is much simpler, the nearest neighbor two-point correlations along
the axes of the isotropic finite system.

\section{Analytical Results}
\label{AR}

In this section we summarize our analytical results for the
two-dimensional Ising, for the spherical model with $2<d<4$, and
for the Gaussian model. Their derivation for the Ising model is
given in Appendix \ref{aIsing}, while ones for the spherical model
are given in Appendix \ref{aspm}.

\subsection{Two-Dimesnional Ising Model}
\label{S2dIsing}

For the two-dimensional  Ising  model on a square lattice with
geometry $L \times M$ the lattice representation of the stress
tensor is well known for a long time \cite{KC71,BK98}. In our
notations, using Eqs. (\ref{st}) and (\ref{xip0p0}), we obtain
\begin{eqnarray}
\label{theconstantIsing}
\lefteqn{\beta
J'(1)\left[\left.\frac{d}{d\lambda} \left(
\frac{\xi_{\parallel,0}(\lambda)}{\xi_{\perp,0}(\lambda)}\right)
\right|_{\lambda=0}\right]^{-1}=} \nonumber \\
& & \frac{\beta J(1+\sinh(2\beta J ))}{2(\sinh(2\beta J)-2\beta J
\cosh(2\beta J)-1)}.
\end{eqnarray}
At the critical point $\beta_c$ of the isotropic system one has
\cite{CW73}
\begin{equation}\label{Isnigcp}
    1=\sinh(2\beta_c J),
\end{equation}
and the right-hand side of the above equation simplifies
essentially becoming simply $-1/(2\sqrt{2})$. Therefore, at
$T=T_c$, the stress tensor reads
\begin{equation}
t_{x,x}(i,j)=\frac{1}{2\sqrt{2}}(S_{i,j}S_{i,j+1}-S_{i,j}S_{i+1,j}),
\end{equation}
which is exactly the form considered in \cite{BK98}. In the limit
$M\rightarrow\infty$ at the critical point $T_c$ of the bulk
system one has \cite{BK98}
\begin{equation}
\label{Isingcritcas}
\langle t_{x,x} \rangle=-\frac{\pi}{6}c
L^{-2},
\end{equation}
where $c=1/2$ is the so-called central charge of the Ising model
\cite{Ca87}. The Casimir amplitude is \cite{Ca87}
\begin{equation}
\label{deltaIsing} \Delta=-\frac{\pi}{6}\ c, \ c=\frac{1}{2}.
\end{equation}
It is easy to see that close to $T_c$ the right-hand side of Eq.
(\ref{theconstantIsing}) becomes
\begin{equation}\label{closetotc}
-\frac{1}{2\sqrt{2}}(1+\frac{\beta-\beta_c}{\beta_c})+O\left((\beta-\beta_c)^2\right).
\end{equation}
Since $\nu=1$ for $2d$ Ising model, it is clear from Eq.
(\ref{Isingcritcas}) that the contributions to the Casimir force
due to the term proportional to $\beta-\beta_c$ in the above
expression will be of the order of $L^{-3}$. Such contributions
will be neglected. Therefore, in the critical region of the finite
system we conclude, that the stress tensor is given by
\begin{eqnarray}\label{stcrIsing}
t_{x,x}(i,j)&=&\frac{1}{2\sqrt{2}}(S_{i,j}S_{i,j+1}-S_{i,j}S_{i+1,j})\nonumber \\
&&+ \frac{1}{2}(\beta_c-\beta)({\cal \hat{H}}-{\cal \hat{H}}_b).
\end{eqnarray}

One can interpret the variance of the stress tensor $\Delta
t_{x,x}(i,j)$ as a variance of a local measurement of the Casimir
force made near the point $(i,j)$. For the leading behavior of the
variance at $T_c$ one then has (see Eq. (\ref{avarsingleIsing}))
\begin{equation}
\Delta t_{x,x}(i,j)\simeq 1-2/\pi.
\end{equation}
Definitely, in addition from the above nonuniversal part the
variance contains also universal parts that  are negligible in
comparison with the nonuniversal one.

As we said above, we will interpret  $\langle t_{x,x}(i,j)
\rangle$ as a local measurement of the Casimir force made near the
point $(i,j)$. Let us imagine, that we are collecting measurements
from all the points belonging to the "surface" $(1,j)$, $j\in
[1,\cdots, M]$ (in the very same way one can consider the opposite
"surface" $(L,j)$, $j\in [1,\cdots, M]$). The surfaces are
important because they are the only place where in an experiment
the Casimir force is experimentally accessible. To characterize
the force measured on the  whole surface, instead of
$t_{x,x}(i,j)$, one has to consider $\sum_{j}t_{x,x}(1,j)$.
Taking, in a first approximation, any local measurement to be
independent from the other ones, one obtains that
\begin{equation}
\Delta\sum_{j}t_{x,x}(1,j) \sim M \Delta t_{x,x}(1,1)\simeq 0.363
\ M, \label{Isingvarianceapprox}
\end{equation}
which implies that, indeed, in agreement with \cite{BAFG02},
\begin{equation}\label{vfIM}
    (\Delta \beta F_C^t)^2 \propto
    N_\parallel^{d-1}=(L_\parallel/a)^{d-1}, \mbox{where} \ \ d=2.
\end{equation}
An estimation can be also derived for
$\Delta\sum_{i,j}t_{x,x}(i,j)$. With a variance of such a type one
deals when, say, Monte Carlo simulations of the force are
performed. One obtains (see Eq. (\ref{avstIsingfinal}))
\begin{equation}\label{vstIsingfinal}
\Delta\sum_{i,j}t_{x,x}(i,j)=
\frac{1}{2}\left[-\frac{1}{2}+\frac{2}{\pi}\right] M L \simeq
0.068 M L. 
\end{equation}
We again observe that the variance of the sum of $t_{x,x}(i,j)$ is
proportional to the total number of summands  in this sum. The
coefficient of proportionality for 2d Ising model, when the sum is
over all spins in the finite system, turns out to be $0.068$.

Unfortunately, as far as we are aware, the finite-size properties
of the free energy of the two-dimensional anisotropic Ising model
under periodic boundary conditions are not available for $T \ne
T_c$. This makes the comparison of the direct derivation of the
force as a derivative of the finite-size scaling excess free
energy and as average of the operator (\ref{st}) a challenging
task. Even more - the behavior of the finite-size free energy of
the isotropic system is only known for moderate values of the
scaling arguments \cite{FF69}. Nevertheless, from \cite{FF69} one
can extract the following results for the scaling functions of the
excess free energy and the Casimir force
\begin{itemize}

\item{excess free energy}

The scaling function of the excess free energy is \cite{BDT00}
\begin{equation}\label{efeIsing}
X_{{\rm ex}}=-\frac{\pi}{12}-\pi \sum_{i=2}^{\infty}\left(
\begin{array}{c}
1/2 \\ i
\end{array}\right) \left(
\frac{x}{2\pi}\right)^{2i}(1-2^{-2i+1})\zeta(2i-1),
\end{equation}
where $-\pi<x<2\pi$, and the scaling variable is $x=8K_c t L$.

\item{Casimir force}

The scaling function of the Casimir force $X_{{\rm Casimir}}$ is
related to that one of the excess free energy via
\begin{equation}\label{XexXcas}
X_{{\rm Casimir}}=X_{{\rm ex}}(x)-x \frac{\partial}{\partial x}
X_{{\rm ex}}(x).
\end{equation}
Then, from (\ref{efeIsing}), one immediately obtains
\begin{eqnarray}\label{efeI}
X_{{\rm Casimir}}&=&-\frac{\pi}{12}-\pi \sum_{i=2}^{\infty}\left(
\begin{array}{c}
1/2 \\ i
\end{array}\right)  \\
& & \times \left(
\frac{x}{2\pi}\right)^{2i}(1-2i)(1-2^{-2i+1})\zeta(2i-1).
\nonumber
\end{eqnarray}
\end{itemize}

\subsection{The Spherical Model}
\label{SM}

We consider a spherical model on a $d$-dimensional hypercubic
lattice $\Lambda \in Z^d$, where $\Lambda=L_1\times L_2\times
\cdots L_d$. Let $L_i=N_i a_i, i=1,\cdots,d$, where $N_i$ is the
number of spins and $a_i$ is the lattice constant along the axis
${\bf e}_i$ with ${\bf e}_i$ being a unit vector along that axis.
With each lattice site $\bf{r}$ one associates a real-valued spin
variable $S_{\bf r}$ which obey the constraint
\begin{equation}
\label{constraint}
\sum_{{\bf r}\in \Lambda} \langle S_{\bf r}^2
\rangle=N,
\end{equation}
where $N=N_1 N_2 \cdots N_d$ is the total number of spins in the
system. The average in (\ref{constraint}) is with respect to the
Hamiltonian of the model which is
\begin{equation}
\label{Hamsm}
\beta{\cal H}=-\frac{1}{2}\beta\sum_{{\bf r}, {\bf
r}'}S_{\bf r}J({\bf r}, {\bf r}')S_{{\bf r}'}+s\sum_{{\bf
r}}S_{{\bf r}}^2.
\end{equation}
In the current article we will consider only the case of
nearest-neighbor interactions, i.e. we take $J({\bf r}, {\bf
r}')=J(|{\bf r}-{\bf r}')|=\ J_j$, if ${\bf r}-{\bf r}'=\pm {\bf
e}_j$, $i=1,\cdots,d$, and $J({\bf r}, {\bf r}')=0$ otherwise.

For such a model it can be shown \cite{BDT00} that, under periodic
boundary conditions, the free energy of the model (per unit spin)
is given by
\begin{eqnarray}
\label{freeenergyspmdef}
\beta f(K,{\bf
N})&=&\frac{1}{2}\left[\ln{\frac{K}{2\pi}}-K\right]+\sup_{w>0}
\left\{ -\frac{1}{2}K w \right.
 \\
& &
 \left. + \frac{1}{2N} \sum_{{\bf k}\in
{\cal B}_\Lambda} \ln{\left[ w+1-\frac{\hat{J}({\bf
k})}{\hat{J}({\bf 0 })}\right]} \right\}, \nonumber
\end{eqnarray}
while the two-point correlation function is
\begin{equation}
\label{cordef}
 G({\bf r},K,{\bf N})=\frac{1}{K N}\sum_{{\bf k}\in
{\cal B}_\Lambda}\frac{e^{i{\bf k}.{\bf r}}}{w+1-\hat{J}({\bf
k})/\hat{J}({\bf 0 })}.
\end{equation}
Here $s=K(w+1)/2$, $K=\beta \hat{J}({\bf 0})$, where $\hat{J}({\bf
k})$ is given by the Fourier transform of the interaction
\begin{equation}
\hat{J}({\bf k})=\sum_{\bf r} J({\bf r})e^{i {\bf k}.{\bf r}},
\end{equation}
and the wave vector ${\bf k}=\{k_1, k_2, \cdots, k_d\}\in {\cal
B}_\Lambda$ is with components  $k_i=2\pi n_i/L_i$, where
$n_i=1,\cdots,N_i$, $i=1,\cdots,d$. The equation for the spherical
field $w$ reads
\begin{equation}
\frac{1}{N}\sum_{{\bf k}\in {\cal
B}_\Lambda}\frac{1}{w+1-\hat{J}({\bf k})/\hat{J}({\bf 0 })}=K,
\end{equation}
which leads immediately to $G({\bf 0},t,{\bf N})=1$.

For nearest neighbor anisotropic interactions it can be shown that
(see Eq. (\ref{acorrl}))
\begin{equation}\label{corrl}
    \frac{\xi_j}{\xi_i}=\sqrt{\frac{b_j}{b_i}}=\sqrt{\frac{J_j}{J_i}},
\end{equation}
where $\xi_j$ is the correlation length  in direction $j$, and
$b_j=J_j/\sum_{j=1}^d J_j$, which leads to the following explicit
form of the stress tensor within the spherical model (see Eq.
(\ref{astsm}))
\begin{eqnarray}
\label{stsm}
t_{\perp \perp}({\bf R}) &=& \frac{\beta J}{d/2}
\left[\sum_{k=1}^{d-1} S_{\bf R} S_{{\bf R}+{\bf e}_k} - (d-1)
S_{\bf R} S_{{\bf R}+{\bf e}_d}\right]\nonumber \\
&& +\frac{1}{d \; \nu}(\beta_c-\beta)\left({\cal \hat{H}}- {\cal
\hat{H}}_b\right).
\end{eqnarray}
Here ${\cal \hat{H}}$ is the Hamiltonian (normalized per unit
particle) of the finite system, and ${\cal \hat{H}}_b$ is that one
of the infinite system.

\subsubsection{Evaluation of the finite-size
excess free energy of the anisotropic system}

First, one can demonstrate that the critical coupling of the
anisotropic bulk system is (see Eq. (\ref{acritT}))
\begin{equation}\label{critT}
    K_c \equiv 2\beta_c \sum_{j=1}^d J_j=W_d(0|{\bf b})=
    \int_0^\infty dx \prod_{j=1}^d e^{-x b_j}I_0(x b_j).
\end{equation}
Then, close to $K=K_c$, when $2<d<4$, for the scaling form of the
{\it excess} free energy $\beta(f-f_b)$ (per spin) in the limit of
a film geometry $N_1, N_2, \cdots, N_{d-1}\rightarrow \infty$ one
obtains (see Eq. \ref{aexcess})
\begin{eqnarray}\label{excess}
\lefteqn{\beta \left[f(K, N_\perp|{\bf  b})-f_b(K|{\bf b})\right]=
\left\{\frac{1}{4}x_1 (y-y_\infty)\right.}
\\
&& \left. -\frac{1}{2}\frac{\Gamma(-d/2)}{(4\pi)^{d/2}}
\left(\frac{b_\perp}{b_\parallel}\right)^{(d-1)/2}\left(y^{d/2}-y_\infty^{d/2}\right)\right.\nonumber
\\
&&\left. -\frac{2}{(2\pi)^{d/2}}
\left(\frac{b_\perp}{b_\parallel}\right)^{(d-1)/2}
y^{d/4}\sum_{q=1}^{\infty}\frac{K_{d/2}(q\sqrt{y})}{q^{d/2}}\right\}N_\perp^{-d}.
\nonumber
\end{eqnarray}
In the above equation
\begin{equation}
\label{x1} x_1=b_\perp (K_c-K)N_\perp^{1/\nu}, \
\nu=\frac{1}{d-2},
\end{equation}
 is the
temperature scaling variable, $y_\infty=2w_b N_\perp^2/b_\perp$ is
the solution of the bulk spherical field equation (see Eq.
(\ref{asfsfeb}))
\begin{equation}\label{sfsfeb}
    -\frac{1}{2}x_1=\frac{\Gamma(1-d/2)}{(4\pi)^{d/2}}\left(\frac{b_\perp}{b_\parallel}\right)^{(d-1)/2}
    y_\infty^{d/2-1},
\end{equation}
while $y=2wN_\perp^2/b_\perp$ is the solution of the finite-size
spherical field equation (see Eq. (\ref{asfsfefinite}))
\begin{eqnarray}\label{sfsfefinite}
   \lefteqn{ -\frac{1}{2}x_1 = \frac{\Gamma(1-d/2)}{(4\pi)^{d/2}}\left(\frac{b_\perp}{b_\parallel}\right)^{(d-1)/2}
    y^{d/2-1}}\\
    && + \frac{2}{(2\pi)^{d/2}}
\left(\frac{b_\perp}{b_\parallel}\right)^{(d-1)/2}
y^{d/4-1/2}\sum_{q=1}^{\infty}\frac{K_{d/2-1}(q\sqrt{y})}{q^{d/2-1}}.
\nonumber
\end{eqnarray}
For the Casimir force one derives (see Eq. (\ref{acassm}))
\begin{eqnarray}
\label{cassm}
 \lefteqn{\beta F_{{\rm Casimir}} = N_\perp^{-d}\left\{ \frac{1}{4}x_1 (y-y_\infty)\right.} \\
 && \left.-(d-1)\left(\frac{b_\perp}{b_\parallel}\right)^{(d-1)/2}\left[\frac{1}{2}\frac{\Gamma(-d/2)}{(4\pi)^{d/2}}
\left(y^{d/2}-y_\infty^{d/2}\right)\right. \right. \nonumber \\
  &&\left. \left. + \frac{2}{(2\pi)^{d/2}}
y^{d/4}\sum_{q=1}^{\infty}\frac{K_{d/2}(q\sqrt{y})}{q^{d/2}}\right]
\right\}, \nonumber
\end{eqnarray}
whereas  for the Casmir amplitudes $\Delta$  we derive (see Eq.
(\ref{adeltasm}))
\begin{equation}\label{deltasm}
    \Delta=-\frac{2}{d(2\pi)^{d/2}}y_c^{d/4+1/2}\sum_{q=1}^{\infty}\frac{K_{d/2+1}(q\sqrt{y_c})}{q^{d/2-1}},
\end{equation}
with $\beta_cF_{{\rm Casimir}}(K_c,L_\perp)=(d-1)\ \Delta \
L_\perp^{-d}$. The exact value of $y_c$ and $\Delta$ in an
explicit form is only known for $d=3$. Then
\begin{equation}
\label{yc} y_c=4\ln^2[(1+\sqrt{5})/2]
\end{equation}
(this value is well-known and seems that has been derived for the
first time in \cite{P72}), and, then, one obtains \cite{D98}
\begin{equation}
\label{y3}
\Delta=-\frac{2\zeta(3)}{5\pi}.
\end{equation}
This is the only exactly know Casimir amplitude for a three
dimensional system.

\subsubsection{Evaluation of the average value of the stress tensor}

For the scaling form of the average value of the stress tensor one
derives (see Eq. (\ref{astscalingformeq}))
\begin{eqnarray}\label{stscalingformeq}
     \langle t_{\perp \perp}({\bf R}) \rangle &=& -\left\{
\frac{2(d-1)}{(2\pi)^{d/2}}\left[y^{d/4}\sum_{q=1}^{\infty}\frac{K_{d/2}(q\sqrt{y})}{q^{d/2}}\right.\right.
\nonumber
\\
& & \left.\left.
+\frac{1}{d}y^{(d+2)/4}\sum_{q=1}^{\infty}\frac{K_{d/2-1}(q\sqrt{y})}{q^{d/2-1}}
\right]\right.\nonumber \\
&& \left.+\frac{(d-2)}{4 d}x_1 (y-y_\infty)\right\}N_{\perp}^{-d}.
\end{eqnarray}
It is also possible to demonstrate (see Appendix \ref{aspm}) that
the above expression is equivalent to $\beta F_{{\rm Casimir}}$
given by Eq. (\ref{cassm}) for the isotropic system (when
$b_\|=b_\perp$), i.e. indeed,
\begin{equation}\label{smfinal}
\langle t_{\perp \perp}({\bf R}) \rangle=\beta F_{{\rm Casimir}}
\end{equation}
for the spherical model.

\subsubsection{Evaluation of the variance of the stress tensor}

For the variance of the Casimir force in the spherical model at
$T=T_c$ one obtains (see Eq. (\ref{vsmnumerical}))
\begin{equation}\label{vsmnumerical}
\Delta t_
{\perp,\perp} \equiv \Delta \sum_{{\bf R}\in
\Lambda}t_{\perp,\perp}({\bf R}) \simeq 0.107\  N_\perp
N_\parallel^2.
\end{equation}
This will be also the leading result everywhere in the critical
region - since it is coming from the nonsingular part of the free
energy around $T=T_c$ an analytical expansion should be possible.
It is also clear that if the summation over ${\bf R}$ in
(\ref{vsmnumerical}) is not over the total number of particles in
$\Lambda$, which is $N_\perp N_\parallel^{d-1}$, but over, say,
all the spins from one of the boundary, then the corresponding
variance will be proportional to the total number of spins there,
i.e. that
\begin{equation}\label{vfsm}
    (\Delta \beta F_C^t)^2 \propto
    N_\parallel^{d-1}=(L_\parallel/a)^{d-1},
\end{equation}
exactly as it has been found in \cite{BAFG02}, see Eq.
(\ref{var}).

\subsection{The Gaussian Model}
\label{GM}

In order to simplify the notations we define the Gaussian model in
the same way that we have defined the spherical model. Actually
the spherical model is a Gaussian type model with one additional
constraint, given by Eq. (\ref{constraint}) fixing the average
length of all the spins in the system. To be more precise, we
suppose that the Hamiltonian ${\cal H}$ of the model is again
given by Eq. (\ref{Hamsm}), where, as before, $s=K(1+w)/2$, and
$K=\beta \tilde{J}(\bf 0)$. The only difference is that now Eq.
(\ref{constraint}) is missing and $w$ is not a quantity which
behavior has to be derived from it, but a parameter which
describes the deviation from the critical point, i.e.
$w=(\beta_c-\beta)/\beta$. As a result, the free energy density of
the model becomes
\begin{equation}\label{freeenergygm}
\beta f(K,{\bf
N})=\frac{1}{2}\left[\ln{\frac{K}{2\pi}}-K\right]+U(w,{\bf
N})-\frac{1}{2}Kw,
\end{equation}
where $U(w,{\bf N})$ is given by Eq. (\ref{u}), while the
two-point correlation function is still determined by Eq.
(\ref{cordef}). Then, for a system with anisotropic short-range
interaction, proceeding in the same way as in Section \ref{SM}, we
derive that Eqs. (\ref{cor}) - (\ref{der}) are still valid,
wherefrom we conclude that in the Gaussian model the stress tensor
again is
\begin{eqnarray}
\label{stGM} t_{\perp \perp}({\bf R}) &=& \frac{\beta J}{d/2}
\left[\sum_{k=1}^{d-1} S_{\bf R} S_{{\bf R}+{\bf e}_k} - (d-1)
S_{\bf R} S_{{\bf R}+{\bf e}_d}\right]\nonumber \\
&& +\frac{1}{d \; \nu}(\beta_c-\beta)\left({\cal \hat{H}} -{\cal
\hat{H}}_b\right),
\end{eqnarray}
where ${\cal \hat{H}}$ is the Hamiltonian (normalized per unit
particle) of the finite and ${\cal \hat{H}}_b$ of the infinite
model.

\subsubsection{Evaluation of the finite-size excess free energy of the anisotropic system}

The analysis of the excess free energy cam be performed along the
same lines as in the case of a spherical model. For example, for
$U(w,{\bf N})$ the Eqs. (\ref{udec})-(\ref{deltauscaling}) and
(\ref{ubfinal}) are still valid. On the basis of these
equations we immediately obtain that in the case of a Gaussian
model the excess free energy in a film geometry is
\begin{eqnarray}\label{excessGM}
\lefteqn{\beta \left[f(K, N_\perp|{\bf  b})-f_b(K|{\bf b})\right]=}\\
&& -\frac{2}{(2\pi)^{d/2}}
\left(\frac{b_\perp}{b_\parallel}\right)^{(d-1)/2}
y^{d/4}\sum_{q=1}^{\infty}\frac{K_{d/2}(q\sqrt{y})}{q^{d/2}}
N_\perp^{-d}. \nonumber
\end{eqnarray}
In the above equation the temperature scaling variable is
\begin{equation}
\label{svGM} y=2 w N_\perp^{1/\nu}/b_\perp,\ \mbox{where}\
\nu=1/2, \ \mbox{and} \ w=\beta_c/\beta-1.
\end{equation}
From Eq. (\ref{excessGM}), using the property of the $K_\nu(x)$
functions \cite{GR73} that
\begin{equation}\label{Kproperties}
\frac{\partial}{\partial y}\left[y^{\nu}K_\nu(a y)\right]=-ay^\nu
K_{\nu-1}(a y),
\end{equation}
we immediately derive that the Casimir force in such an
anisotropic Gaussian model is given by
\begin{eqnarray}
\label{casGM}
 \beta F_{{\rm Casimir}} &=& -\frac{2}{(2\pi)^{d/2}}\left(\frac{b_\perp}{b_\parallel}\right)^{(d-1)/2}N_\perp^{-d} \\
  && \times\left\{(d-1) y^{d/4}\sum_{q=1}^{\infty}\frac{K_{d/2}(q\sqrt{y})}{q^{d/2}}\right. \nonumber \\
  && \left.+
  y^{d/4+1/2}\sum_{q=1}^{\infty}\frac{K_{d/2-1}(q\sqrt{y})}{q^{d/2-1}}\right\}.
  \nonumber
\end{eqnarray}
Note that despite the similarities with the spherical model both
the excess free energy and the Casimir force differs essentially
for the two models. Let us demonstrate that even more explicitly
on the example of the Casimir amplitudes $\Delta$. We remind that
in he spherical model they are given, for $2<d<4$, by Eq.
(\ref{deltasm}) where $y_c$ is the solution of the spherical field
equation at $\beta=\beta_c$. In explicit form we have been able to
solve this equation and to calculate $\Delta$ only for $d=3$. The
situation with the Gaussian model is much simpler. At the critical
point $\beta=\beta_c$ one has $y=0$, and, therefore, from Eq.
(\ref{casGM}), or Eq. (\ref{excessGM}), we obtain (in the
isotropic system)
\begin{equation}\label{deltaGM}
    \Delta=-\frac{\Gamma(d/2)\zeta(d)}{\pi^{d/2}}.
\end{equation}
So, for $d=3$ one has $\Delta=-\zeta(3)/2\pi$ and, therefore,
\begin{equation}\label{relationSMGM}
    \Delta_{{\rm Spherical \ Model}}=\frac{4}{5} \Delta_{{\rm Gaussian \
    Model}}, \ d=3.
\end{equation}

\subsubsection{Evaluation of the average value of the stress tensor}

Having in mind that $u=\frac{\partial}{\partial \beta}(\beta f)$
and using (\ref{Kproperties}), for the difference of the
finite-size and bulk internal energy densities one can easily
derive from (\ref{freeenergygm}) and (\ref{excessGM})
\begin{equation}\label{uGM}
u-u_b=-\frac{\beta_c/\beta}{(2\pi)^{d/2}}
\frac{y^{d/4+1/2}}{(\beta_c-\beta)}\sum_{q=1}^{\infty}\frac{K_{d/2-1}(q\sqrt{y})}{q^{d/2-1}}N_\perp^{-d},
\end{equation}
where $y=2dwN_\perp^2$ and $w=\beta_c/\beta-1$. Next, from
(\ref{cordef}), or (\ref{freeenergygm}), for the stress tensor
(\ref{stGM}) of the Gaussian model we derive that
\begin{eqnarray}\label{straverageGM}
\lefteqn{\langle t_{\perp \perp}({\bf R}) \rangle =
\frac{d-1}{d}\frac{1}{N}\sum_{{\bf k}\in {\cal
B}_{\Lambda}}\frac{\cos(k_1 a_1)-\cos(k_d
a_d)}{d(1+w)-\sum_{j=1}^d \cos(k_j a_j)}} \nonumber \\
&& -\frac{\beta_c/\beta}{d}\frac{2}{(2\pi)^{d/2}}
y^{d/4+1/2}\sum_{q=1}^{\infty}\frac{K_{d/2-1}(q\sqrt{y})}{q^{d/2-1}}N_\perp^{-d},
\ \ \ \ \ \ \ \ \
\end{eqnarray}
where we have taken into account that $\nu=1/2$. Applying to the
first row in the above equation the same way of acting, as in the
case of the spherical model, and replacing $\beta_c/\beta$ by $1$
(since we are close to the critical point), we derive
\begin{eqnarray}\label{stscalingformGM}
    \lefteqn{ \langle t_{\perp \perp}({\bf R}) \rangle = -\frac{2}{d(2\pi)^{d/2}}\left\{
y^{d/4+1/2}\sum_{q=1}^{\infty}\frac{K_{d/2-1}(q\sqrt{y})}{q^{d/2-1}}
\right.} \nonumber
\\
& &\left.
+(d-1)y^{(d+2)/4}\sum_{q=1}^{\infty}\frac{K_{1+d/2}(q\sqrt{y})}{q^{d/2-1}}
\right\}N_\perp^{-d}.\ \ \ \ \
\end{eqnarray}
Now it only remains to show that the right-hand side of the above
equation is indeed equal to the right-hand side of Eq.
(\ref{casGM}) (for $b_\perp=b_\parallel$), which gives the Casimir
force calculated in a direct manner as a derivative of the
finite-size free energy with respect  of the size of the system.
In order to demonstrate this, let us note that, according to the
identity (\ref{Kidentity}),
\begin{equation}\label{Kpr}
K_{d/2+1}(x)=K_{d/2-1}(x)+\frac{d}{x} K_{d/2}(x).
\end{equation}
Inserting (\ref{Kpr}) in (\ref{stscalingformGM}) and comparing the
result with Eq. (\ref{casGM}), we conclude that, indeed,
\begin{equation}\label{gmfinal}
\langle t_{\perp \perp}({\bf R}) \rangle=\beta F_{{\rm Casimir}}
\end{equation}
for the Gaussian model.

\subsubsection{Evaluation of the variance of the stress tensor}

For the variance of the stress tensor all the equations from the
corresponding part of Appendix \ref{aspm} for the spherical model
are still valid. That is because the leading contribution of the
variance is stemming from the regular part of the bulk free energy
$U_b(0|{\bf b})$ (see Eq. (\ref{ub0})) evaluated at $T=T_c$ (see
Eq. (\ref{vstsm})). This observation leads to the conclusion that,
as in the spherical model case,
\begin{equation}\label{vsmnumericalGM}
\Delta t_{\perp,\perp} \equiv
\Delta \sum_{{\bf R}\in \Lambda}t_{\perp,\perp}({\bf R}) \simeq
0.107\  N_\perp N_\parallel^2,
\end{equation}
where the summation is over all the spins of the system.

\section{Monte-Carlo Results}
\label{SMCR}

The foundation of our Monte-Carlo investigations of the critical
Casimir force is laid by Eqs.(\ref{st}) and (\ref{FCasimir}), respectively.
Apart from the a priory unknown coefficient $J'(1)/R'(0)$ and the bulk
energy density $u_b$ the numerical evaluation of Eq.(\ref{st}) is
absolutely straightforward and apart from usual algorithmic precautions
in the critical regime no special techniques are required. However,
as has become obvious in, e.g., Eq.(\ref{vstIsingfinal}), the statistical
error of the estimate will increase with the system size if the number
of Monte-Carlo sweeps is kept constant. In order to approach the asymptotic
regime larger system sizes, say, $N_\parallel = 120$ and $N_\perp = 20$
lattice sites in $d=3$ are required which means that a reliable estimation
of the Casimir force remains computationally demanding as far as the required
CPU time is concerned.

We employ a hybrid algorithm \cite{PFL02} which consists of Metropolis
\cite{Met53} and single cluster updates \cite{Wolff89}
for the Ising model, for XY and Heisenberg simulations overrelaxation
updates \cite{KFL93} are employed as a third update method. Cluster updates
are only used in the immediate vicinity of the critical point, e.g., for
$-0.02 \leq t \leq 0.02$ for the system size indicated above. Typically,
we have performed between $4.8 \times 10^6$ and $9.6 \times 10^6$
Monte-Carlo steps per spin. In order
to cope with the high demand of CPU time for larger systems we have performed
part of our simulation in parallel on two-processor Intel Xeon system and
on a four-processor DEC Alpha system using the OpenMP Standard for SMP
programming. A few runs have also been performed on a two-processor AMD
Opteron system.

We first investigated the energy dependent contribution $(\beta_c(0)-\beta)
(u - u_b)$ to Eqs.(\ref{st}) and (\ref{FCasimir}) by a series of simulations
on a cubic geometry for $N_\parallel = N_\perp = 20 \dots 80$ in order to
obtain reliable estimates for the bulk energy density $u_b$. It turns
out, that within the range $-0.2 \leq t \leq 0.2$ of reduced
temperatures, various aspect ratios $N_\parallel / N_\perp = 3, 4, 6, 8$,
and several system sizes $N_\parallel = 60 \dots 120$ the energy dependent
contribution $(\beta_c(0)-\beta) (u - u_b)/d\nu$ is always negligible. As
a typical result we obtained that for forces of the order $10^{-1}$ with
a statistical error in the range $10^{-2} \dots 10^{-3}$ the energy
contribution $(\beta_c(0)-\beta) (u - u_b)/d\nu$ remains in the range
$10^{-4} \dots 10^{-5}$ for all models. The prefactor $J'(1) / R'(0)$
roughly evaluates to $J'(1) / R'(0) \simeq 0.3$ in all cases. We therefore
conlude that we can safely ignore the energy contribution to the Casimir
force for our Monte-Carlo investigations of the Ising, the XY, and the
Heisenberg model in three dimensions.

\subsection{Three-Dimensional Ising Model}
\label{S3dIsing}

\begin{figure}[h]
\includegraphics[scale=0.42]{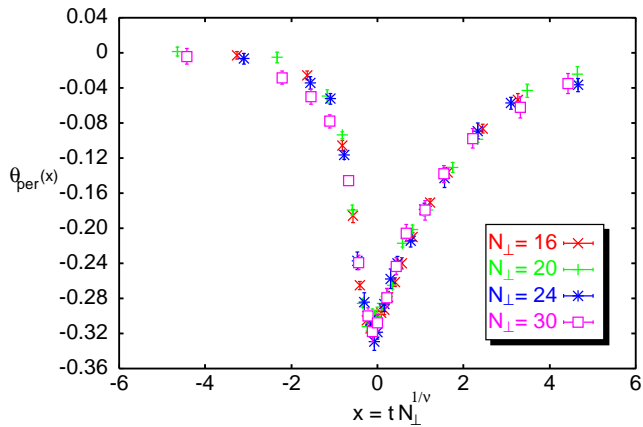}
\caption{Scaling function $\theta_{per}(x)$ of the Casimir force
for the $d=3$ Ising model in a slab geometry for periodic boundary
conditions as function of the scaling variable $x=t
N_\perp^{1/\nu}$, where $N_\perp = L_\perp/a$ is the number of
lattice layers. The aspect ratio is chosen as $N_\parallel /
N_\perp = 6$ (see main text). Finite size scaling according to our
expectation is confirmed within two standard deviations, where
$\nu = 0.63$ has been chosen. The vertical scale has been adjusted
according to the estimate $\theta_{per}(0) \equiv 2 \Delta_{per,n=1}
= -0.306$ (see Ref.\protect\cite{MK97}). The error bars displayed
here correspond to one standard deviation. \label{IsTzzscal}}
\end{figure}

As expounded above, we have neglected the energy dependent contribution
to Eq.(\ref{FCasimir}) for our Monte-Carlo evaluations of the scaling
Function $\theta_{per}(x)$, $x = t (L_\perp/a)^{1/\nu}$ of the Casimir force.
From extended simulations for various aspect ratios $N_\parallel/N_\perp =
3, 4, 6$, and $8$ we have arrived at the conclusion that corrections
to $\theta_{per}(x)$ due to finite aspect ratio are by far negligible
within the statistical error for $N_\parallel/N_\perp = 6$. In fact,
our results for $N_\parallel/N_\perp = 4$ can hardly be distinguished from
corresponding results for larger aspect ratios. We have therefore fixed
the aspect ratio to the value 6 and performed simulations for $N_\perp =
16, 20, 24$, and $30$. The resulting scaling plot for $\theta_{per}(x)$
is shown in Fig.\ref{IsTzzscal}.
For $T \geq T_c$ finite-size scaling works very well, whereas for $T < T_c$
data collapse for $N_\perp = 30$ is not as good. However, the data collapse
improves upon increasing the statistics and so we have performed
$9.6 \times 10^6$ Monte-Carlo steps per lattice site for the largest
lattice $N_\perp = 30$ for $T < T_c$. With the estimate $\nu = 0.63$
for the correlation length exponent we finally obtain scaling within
two standard deviations. The scaling function $\theta_{per}(x)$ decays
exponentially for $x \to \pm \infty$ and has its minimum
{\em below} $T_c$. However, due to the magnitude of the statistical
error its location cannot be determined accurately enough to exclude
$x = 0$ with reasonable certainty. The Monte-Carlo data for the Casimir
force are not normalized due to the a priory unknown prefactor
$J'(1) / R'(0)$ in Eq.(\ref{FCasimir}). The data displayed in
Fig.\ref{IsTzzscal} have therefore been scaled in such a way that
$\theta_{per}(0) \equiv 2 \Delta_{per,n=1}$ is given by the best known
estimate $\Delta_{per,n=1} \simeq -0.153$ for the Casimir anmplitude
$\Delta_{per,n=1}$ for the three-dimensional Ising model \cite{MK97}.

\begin{figure}[h]
\includegraphics[scale=0.42]{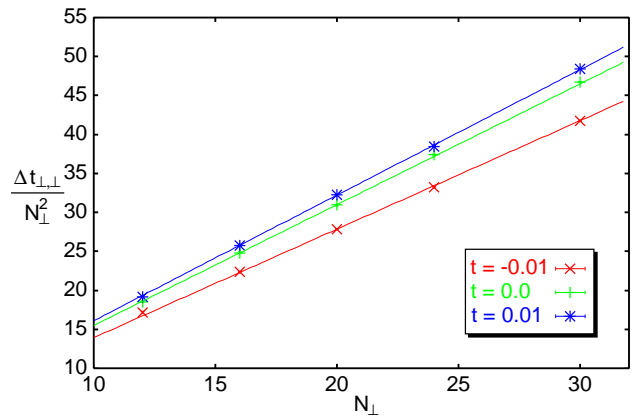}
\caption{Variance $\Delta t_{\perp,\perp}$ of the stress tensor
for the Ising model in $d = 3$
(see Eqs.(\protect\ref{vsmnumerical}) and (\protect\ref{vsmnumericalGM})),
normalized to $N_\perp^2$ at fixed aspect ratio $N_\parallel / N_\perp = 6$
as function of $N_\perp$ for different reduced temperatures $t$ in the
critical regime. The behavior is linear as indicated by the straight lines
connecting the data points. Their slopes have been evaluated as 1.39
for $t = -0.01$, 1.55 for $t = 0$, and 1.61 for $t = 0.01$. The statistical
error of the data (one standard deviation) is smaller than the symbol size.
\label{IsDeltat}}
\end{figure}

The scaling function displayed in Fig.\ref{IsTzzscal} has been obtained
from Eq.(\ref{FCasimir}), where a spatial average over all lattice sites
is performed. As expounded in Sec.III (see also Ref.\cite{BAFG02}) this
leads to a certain size dependence of the {\em variance} of the stress
tensor as, e.g., given by Eq.(\ref{vsmnumerical}) for the spherical
model and by Eq.(\ref{vsmnumericalGM}) for the Gaussian model. In order
to investigate the variance also for the Ising model in $d=3$ we have
recorded the distribution function of the stress tensor during our
Monte-Carlo simulations. It turns out that the shape of the distribution
function is captured by a Gaussian distributiuon to a very high degree of
accuracy also for the Ising model (see Ref.\cite{BAFG02}). We are therefore
able to extract the variance of the stress tensor average from a least
square fit of the measured distribution function to a Gaussian, where
the variance is one of the fit parameters. Guided by Eqs.(\ref{vsmnumerical})
and (\ref{vsmnumericalGM}) we have normalized the variance to $N_\perp^2$
in order to obtain a linear law at fixed aspect ratio. Our results for
$N_\parallel / N_\perp = 6$ are displayed in Fig.\ref{IsDeltat}.
The functional dependence is indeed linear and the slope at $t = 0$
$(T = T_c)$ is 1.55 as compared to $0.107 (N_\parallel/N_\perp)^2 \simeq
3.85$ for $N_\parallel/N_\perp=6$ according to Eqs.(\ref{vsmnumerical})
and (\ref{vsmnumericalGM}) for the spherical and the Gaussian model.
The strict linearity also prevails for other
temperatures in the scaling regime as shown in Fig.\ref{IsDeltat}.
The quadratic dependence of $\Delta t_{\perp,\perp} / N_\perp^2$ on the
aspect ratio has also been confirmed for the Ising model in $d = 3$ at
$T = T_c$ from simulations at different aspect ratios (not shown).

\subsection{Three-Dimensional XY Model}
\label{S3dXYM}

In accordance with our findings for the Ising model we find that the
value 6 for the aspect ratio of the simulation lattice is also a good
choice for the XY model. We have performed simulations for $N_\perp =
16, 20, 24$, and $30$, where $4.8 \times 10^6$ Monte-Carlo steps per
lattice site have been performed for all lattice sizes. It turns out
that the energy dependent contribution to Eq.(\ref{FCasimir}) can again
be disregarded within the statistical error obtained from the simulations.

As in the Ising case we determine the normalization factor $J'(1) / R'(0)$
in Eq.(\ref{FCasimir}) from the requirement $\theta_{per}(0)=2
\Delta_{per,n=2}$.
Unfortunately, all estimates for $\Delta_{per,n=2}$, which are currently
available, are based on the $\varepsilon$-expansions quoted, e.g., in
Ref.\cite{K94}. Independent Monte-Carlo estimates for the Casimir amplitudes
of the XY model do not exist and rigorous results for the two - dimensional
XY model are limited to temperatures below the Kosterlitz - Thouless
Temperature, where the model renormalizes towards the two - dimensional
Gaussian fixed point. The Gaussian model in $d=2$ is characterized by the
central charge $c = 1$ and therefore the Casimir amplitude for the two -
dimensional XY model in the low temperature limit is given by
$\Delta_{per,n=2,d=2} = -\pi c / 6 = -\pi / 6 \simeq -0.5236$
\cite{BCN86}.

\begin{figure}[h]
\includegraphics[scale=0.42]{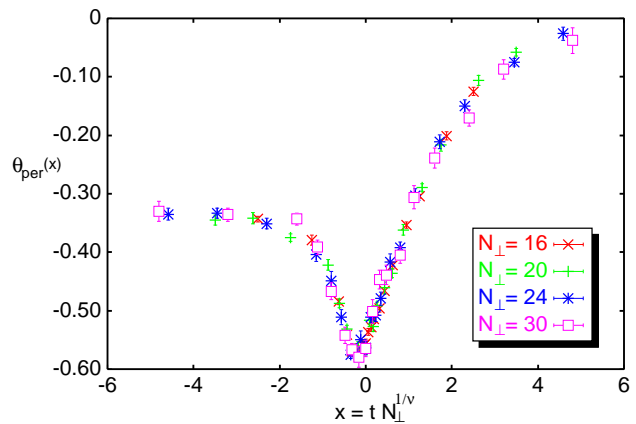}
\caption{Scaling function $\theta_{per}(x)$ of the Casimir force for the
$d=3$ XY model in a slab geometry for periodic boundary conditions
as function of the scaling variable $x=t N_\perp^{1/\nu}$, where
$N_\perp = L_\perp/a$ is the number of lattice layers. The aspect
ratio is chosen as $N_\parallel / N_\perp = 6$ (see main text).
Finite size scaling according to our expectation is confirmed
within two standard deviations, where $\nu = 0.67$ has been chosen.
The vertical scale has been adjusted according to the estimate
$\theta_{per}(0) \equiv 2 \Delta_{per,n=2} = -0.56$ (see main text).
The error bars displayed here correspond to one standard deviation.
\label{XYTzzscal}}
\end{figure}

Apparently, the $\varepsilon$-expansion underestimates
the magnitude of the Casimir amplitude $\Delta_{per,n=1}$ of the critical
Ising model in $d=3$, i.e., $\varepsilon = 1$. From the structure of the
critical Ginzburg - Landau $\phi^4$ theory and the nature of the two - loop
approximation to the Casimir amplitude we expect that the $\varepsilon$ -
expansion will also underestimate the magnitude of $\Delta_{per,n}$ for any
$n$. This leads us to the conclusion that the $\varepsilon$-expansion of the
{\em ratio}
\begin{equation} \label{Dperratio}
\frac{\Delta_{per,n}}{\Delta_{per,n=1}} = n \left[1 - \frac{5}{4}\ \varepsilon
\left(\frac{n+2}{n+8} - \frac{1}{3}\right) + {\cal O}(\varepsilon^2)\right]
\end{equation}
is more accurate in $d=3$ than the $\varepsilon$-expansion for numerator and
denominator individually (see Ref.\cite{K94}). We therefore adopt the
approximation
\begin{equation} \label{Dpern}
\Delta_{per,n} \simeq -0.153\ n \left[1 - \frac{5}{4}
\left(\frac{n+2}{n+8} - \frac{1}{3}\right)\right]
\end{equation}
as our estimate for $\Delta_{per,n}$ in $d=3$ in the following, where
$\Delta_{per,n=1} = -0.153$ (see above) has been used. From Eq.(\ref{Dpern})
we then have
\begin{equation} \label{Dper2}
\Delta_{per,2} \simeq -0.28
\end{equation}
for the three dimensional XY model. The resulting scaling plot for
$\theta_{per}(x)$ is shown in Fig.\ref{XYTzzscal}.
For $T \geq T_c$ finite-size scaling works very well, whereas for $T < T_c$
data collapse for $N_\perp = 30$ is again not as good. However, the data
collapse ist still acceptable within two standard deviations, so we refrain
from performind additional runs here. The scaling function $\theta_{per}(x)$
decays exponentially above $T_c$ for $x \to \infty$ and displays a minimum
below $T_c$. Unlike the Ising model the XY model exhibits lang-ranged
correlations also below $T_c$ (goldstone modes) which are a prominent
feature in Fig.\ref{XYTzzscal}. The scaling function $\theta_{per}
(x \to -\infty)$ saturates at about half its minimum value and does no
longer decay to zero.

\begin{figure}[h]
\includegraphics[scale=0.42]{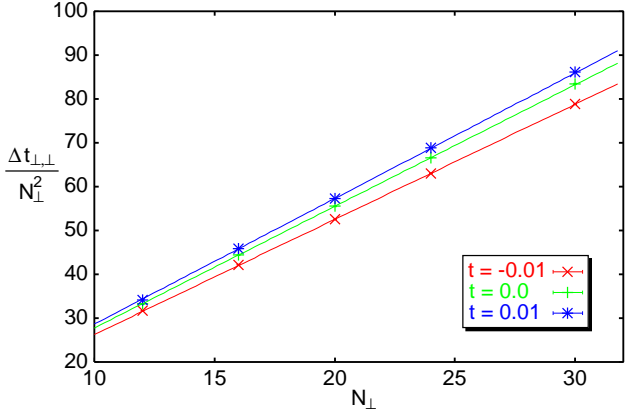}
\caption{Variance $\Delta t_{\perp,\perp}$ of the stress tensor
for the XY model in $d = 3$
(see Eqs.(\protect\ref{vsmnumerical}) and (\protect\ref{vsmnumericalGM})),
normalized to $N_\perp^2$ at fixed aspect ratio $N_\parallel / N_\perp = 6$
as function of $N_\perp$ for different reduced temperatures $t$ in the
critical regime. The behavior is linear as indicated by the straight lines
connecting the data points. Their slopes have been evaluated as 2.63
for $t = -0.01$, 2.78 for $t = 0$, and 2.86 for $t = 0.01$. The statistical
error of the data (one standard deviation) is smaller than the symbol size.
\label{XYDeltat}}
\end{figure}

We have also evaluated the size dependence of the variance of the stress
tensor for the XY model along the lines of the previous analysis
for the Ising model. The distribution function of the stress tensor is
again given by a Gaussian to a very high accuracy. The corresponding
result for $\Delta t_{\perp,\perp}/N_\perp^2$ is shown in Fig.\ref{XYDeltat}.
The functional dependence is again linear and the slope at $t = 0$
$(T = T_c)$ is 2.78 as compared to $0.107 (N_\parallel/N_\perp)^2
\simeq 3.85$ (see previous subsection and Eqs.(\ref{vsmnumerical}) and
(\ref{vsmnumericalGM})) for the spherical and the Gaussian model.
The strict linearity also prevails for other temperatures in the
scaling regime as shown in Fig.\ref{XYDeltat}. In summary the XY
model behaves just as the Ising model with respect to the variance
of the stress tensor.

\subsection{Three-Dimensional Heisenberg Model}
\label{S3dHM}

We have repeated the simulations finally for the Heisenberg model in
$d = 3$ with the same geometric and statistical data as for the XY
model for the same reasons discussed above. We note again that the
energy dependent contribution to Eq.(\ref{FCasimir}) can be disregarded
within the statistical error obtained from the simulations.

\begin{figure}
\includegraphics[scale=0.42]{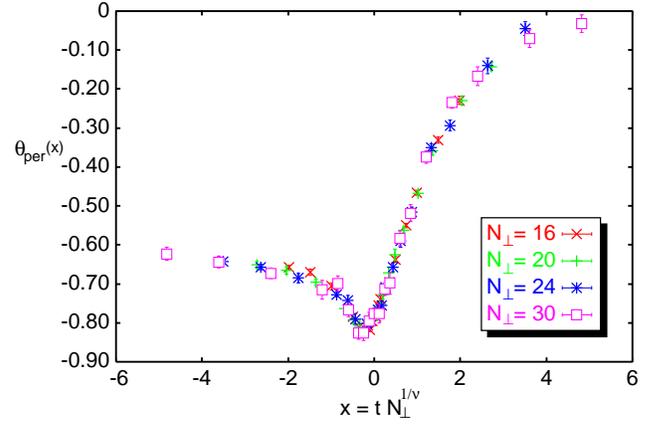}
\caption{Scaling function $\theta_{per}(x)$ of the Casimir force for the
$d=3$ Heisenberg model in a slab geometry for periodic boundary conditions
as function of the scaling variable $x=t N_\perp^{1/\nu}$, where
$N_\perp = L_\perp/a$ is the number of lattice layers. The aspect
ratio is chosen as $N_\parallel / N_\perp = 6$.
Finite size scaling according to our expectation is confirmed
within two standard deviations, where $\nu = 0.71$ has been chosen.
The vertical scale has been adjusted according to the estimate
$\theta_{per}(0) \equiv 2 \Delta_{per,n=3} = -0.78$ (see main text).
The error bars displayed here correspond to one standard deviation.
\label{HTzzscal}}
\end{figure}

The normalization factor $J'(1) / R'(0)$ in Eq.(\ref{FCasimir}) is
determined from the requirement $\theta_{per}(0)=2 \Delta_{per,n=3}$,
where the estimate
\begin{equation} \label{Dper3}
\Delta_{per,3} \simeq -0.39
\end{equation}
used here has been obtained from Eq.(\ref{Dpern}) for $n = 3$.
The resulting scaling plot for $\theta_{per}(x)$ is shown in
Fig.\ref{HTzzscal}.
For $T \geq T_c$ finite-size scaling works very well, whereas for $T < T_c$
the scatter of the date is larger than for the XY model. However, the data
collapse ist still acceptable within two standard deviations.
The qualitative shape of the scaling function $\theta_{per}(x)$ is the
same as for the XY model. The Heisenberg model also exhibits lang-ranged
correlations below $T_c$ (goldstone modes) which are a prominent
feature in Fig.\ref{HTzzscal}. The scaling function $\theta_{per}
(x \to -\infty)$ saturates at about three quarters of its minimum value.

\begin{figure}[h]
\includegraphics[scale=0.42]{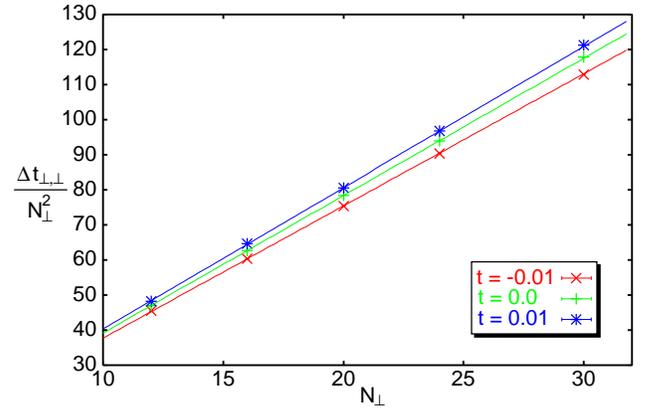}
\caption{Variance $\Delta t_{\perp,\perp}$ of the stress tensor
for the Heisenberg model in $d = 3$
(see Eqs.(\protect\ref{vsmnumerical}) and (\protect\ref{vsmnumericalGM})),
normalized to $N_\perp^2$ at fixed aspect ratio $N_\parallel / N_\perp = 6$
as function of $N_\perp$ for different reduced temperatures $t$ in the
critical regime. The behavior is linear as indicated by the straight lines
connecting the data points. Their slopes have been evaluated as 3.77
for $t = -0.01$, 3.92 for $t = 0$, and 4.03 for $t = 0.01$. The statistical
error of the data (one standard deviation) is smaller than the symbol size.
\label{HDeltat}}
\end{figure}

Finally, have evaluated the size dependence of the variance of the stress
tensor for the Heisenberg model along the lines of the previous analyses
for the Ising and the XY model. As before the distribution function of the
stress tensor is given by a Gaussian to a very high accuracy . The
corresponding result for $\Delta t_{\perp,\perp}/N_\perp^2$ is shown
in Fig.\ref{HDeltat}.
The functional dependence is linear and the slope at $t = 0$
$(T = T_c)$ is 3.92 as compared to $0.107 (N_\parallel/N_\perp)^2
\simeq 3.85$ (see previous subsections and Eqs.(\ref{vsmnumerical}) and
(\ref{vsmnumericalGM})) for the spherical and the Gaussian model.
The strict linearity also prevails for other temperatures in the
scaling regime as shown in Fig.\ref{XYDeltat}. In summary the Heisenberg
model behaves just as the Ising and the XY model with respect to the variance
of the stress tensor.

Apart from different slopes there no specific differences in the behavior of
the variance of the stress tensor for all spin models investigated here in
$d=3$. However, the scaling function of the Casimir force does display
specific differences as one may expect from the presence and the increasing
dominance of Goldstone modes below $T_c$.

\section{Summary and concluding remarks}
\label{SSu}

In the current article an operator - the stress tensor operator -
on a finite lattice systems has been constructed so, that its
average value gives the universal behavior of the thermodynamic
Casimir force near the critical point of a system with
short-ranged interactions (see Eq. (\ref{st})). The definition of
the operator holds in systems in which the hyperscaling is valid
(for $O(n)$ models that are systems with dimensionality $2<d<4$).
Its explicit form for the two-dimensional Ising model is (see Eq.
(\ref{stcrIsing}))
\begin{eqnarray}\label{dstcrIsing}
t_{x,x}(i,j)&=&\frac{1}{2\sqrt{2}}(S_{i,j}S_{i,j+1}-S_{i,j}S_{i+1,j})\nonumber \\
&&+ \frac{1}{2}(\beta_c-\beta)({\cal \hat{H}}-{\cal \hat{H}}_b),
\end{eqnarray}
while that one for the $d$-dimensional ($2<d<4$) spherical and the
$d$-dimensional Gaussian models is (see Eqs. (\ref{stsm}) and
(\ref{stGM}), respectively)
\begin{eqnarray}
\label{dstsm} t_{\perp \perp}({\bf R}) &=& \frac{\beta J}{d/2}
\left[\sum_{k=1}^{d-1} S_{\bf R} S_{{\bf R}+{\bf e}_k} - (d-1)
S_{\bf R} S_{{\bf R}+{\bf e}_d}\right]\nonumber \\
&& +\frac{1}{d \; \nu}(\beta_c-\beta)\left({\cal \hat{H}}- {\cal
\hat{H}}_b\right).
\end{eqnarray}
Here ${\cal \hat{H}}$ is the Hamiltonian (normalized per unit
particle) of the finite system, and ${\cal \hat{H}}_b$ is that one
of the infinite system. For the spherical model one has to take
into account that $\nu=1/(d-2)$, while $\nu=1/2$ for the Gaussian
model. On the example of the two-dimensional Ising model, the
spherical model with $2<d<4$ and the Gaussian model we have
verified via exact calculations the correctness of the above
presentation. They reproduce the correct values of the Casimir
amplitudes at $T=T_c$ and, for the spherical and the Gaussian
models the expressions for the force derived via the excess free
energy and via averaging the stress tensor operator are giving the
same results. The amplitudes and the force near the critical point
of the bulk system turns out, as expected, to be universal and is
in full accordance with the finite-size scaling theory. An
evaluation of the variance of the so defined Casimir force has
been also performed. If the summation is performed over all the
particles within the system the corresponding result for the
two-dimensional Ising model is (see Eq. (\ref{vstIsingfinal}))
\begin{eqnarray}\label{dvstIsingfinal}
\Delta\sum_{i,j}t_{x,x}(i,j)=
\frac{1}{2}\left[-\frac{1}{2}+\frac{2}{\pi}\right] N_\perp
N_\parallel  \simeq 0.068 N_\perp N_\parallel, \ \ \
\end{eqnarray}
while that one for the three-dimensional spherical and the
Gaussian models is (see Eqs. (\ref{vsmnumerical}) and
(\ref{vsmnumericalGM}))
\begin{equation}\label{dvsmnumerical}
\Delta t_ {\perp,\perp} \equiv \Delta \sum_{{\bf R}\in
\Lambda}t_{\perp,\perp}({\bf R}) \simeq 0.107\  N_\perp
N_\parallel^2.
\end{equation}
The average values of the above stress tensor operator are
\begin{equation}
\label{davIsing} \langle \sum_{i,j}t_{x,x}(i,j) \rangle
=-\frac{\pi}{12 N_\perp^2} \ \left(N_\perp N_\parallel \right)
\end{equation}
for the two-dimensional Ising model (see Eq.
(\ref{Isingcritcas})),
\begin{eqnarray}\label{davspm}
\langle t_ {\perp,\perp}\rangle & \equiv & \langle \sum_{{\bf
R}\in \Lambda}t_{\perp,\perp}({\bf R}) \rangle =
-\frac{4\zeta(3)}{5\pi N_\perp^3} \left( N_\perp N_\|^2 \right) \\
 & \simeq &  -\frac{0.306}{N_\perp^3} \ \left( N_\perp N_\parallel^2
\right)
\end{eqnarray}
for the three-dimensional spherical model (see Eq. (\ref{y3})),
and
\begin{eqnarray}\label{davGM}
\langle t_{\perp,\perp}\rangle & \equiv & \langle \sum_{{\bf R}\in
\Lambda}t_{\perp,\perp}({\bf R}) \rangle =
-\frac{\zeta(3)}{2\pi N_\perp^3} \left( N_\perp N_\|^2 \right) \\
 & \simeq &  -\frac{0.382}{N_\perp^3} \ \left( N_\perp N_\parallel^2
\right),
\end{eqnarray}
for the three-dimensional Gaussian model (see Eq.
(\ref{deltaGM})). For the "noise-over-signal" ratio
\begin{equation}\label{defrv}
\rho_V=\frac{\sqrt{\Delta t_{\perp,\perp}}}{\langle
t_{\perp,\perp}\rangle}
\end{equation}
of the so-measured force from the above results one then derives
\begin{equation}\label{rVIsing}
    \rho_V\simeq 0.159 \left( \frac{N_\perp}{N_\|} \right)^{1/2}
    N_\perp,
\end{equation}
for the Ising model,
\begin{equation}\label{rVspm}
    \rho_V \simeq 1.069 \left( \frac{N_\perp}{N_\|} \right)
    N_\perp^{3/2},
\end{equation}
for the spherical model, and
\begin{equation}\label{rVGM}
    \rho_V \simeq 0.856 \left( \frac{N_\perp}{N_\|} \right)
    N_\perp^{3/2},
\end{equation}
for the Gaussian model. In the general case of a $d$-dimensional
critical system the corresponding ratio at the bulk critical point
is
\begin{equation}\label{rVgen}
    \rho_V= \frac{D}{(d-1)\Delta}
    \left( \frac{N_\perp}{N_\|} \right)^{(d-1)/2}   N_\perp^{d/2},
\end{equation}
where $D=D(T \rightarrow T_c)$ is a nonuniversal constant that
describes the behavior of the variance of the tensor, i.e.
\begin{equation}
\label{DstV} \Delta t_{\perp,\perp}\simeq \frac{D^2(T)}{\beta^2}
N_\perp N_\|^{d-1},
\end{equation}
$D(T)$ is a slowly varying nonuniversal function of $T$ close to
$T_c$, and $\Delta$ is the usual Casimir amplitude.

\begin{figure}
\includegraphics[scale=0.42]{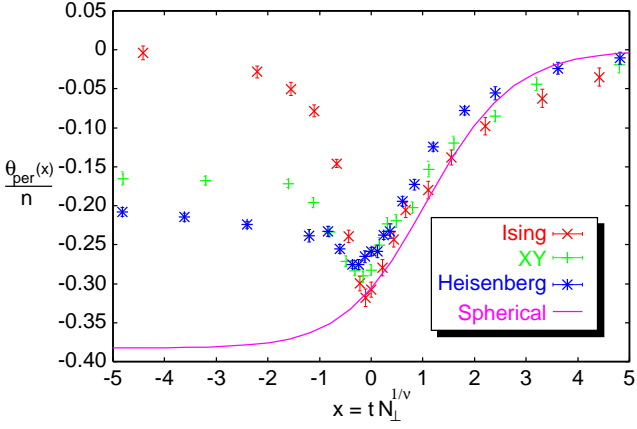}
\caption{Scaling function $\theta_{per}(x)/n$ of the Casimir force
in $d=3$ in a slab geometry for periodic boundary conditions
as function of the scaling variable $x=t N_\perp^{1/\nu}$, where
$N_\perp = L_\perp/a$ is the number of lattice layers. Monte Carlo data
are shown for the Ising model ($+,\ n=1$), the XY model ($\times,\ n=2$)
and the Heisenberg model ($*,\ n=3$) with lattice size $N_\| = 180$
and $N_\perp = 30$. The solid line shows
$\theta_{per}(x)/n$ in the spherical limit ($n \to \infty$).
\label{all}}
\end{figure}

Based on the proposed new operator, Monte Carlo calculations has
been performed and the Casimir force scaling functions has been
determined for the three dimensional Ising, $XY$ and Heisenberg
models. The scaling functions decay exponentially to zero above
the critical temperature. The same happens for the Ising model
also below $T_c$, while for the $XY$ and Heisenberg models
they tend to a constant because of the existence of the Goldstone
modes in this regime in these two models. Our results for $O(n)$
spin models, $n = 1,2,3,\infty$, are summarized in Fig.\ref{all}.
The data for $\theta_{per}(x)$ are normalized to $n$ in order to
obtain a direct comparison with the spherical limit, for which the exact
result is shown.

Our results confirm that one has to take into account the ratio between
the thickness of the film and its lateral dimensions, when planning the
settlement of an experiment, in order to achieve the desired noise-over-signal
ratio. The numerical results that are presented can be considered as a type
of such "measuring" of the force by Monte Carlo methods. They demonstrate
clearly, that high accuracy in such type of measurement of the force is
indeed possible to achieve. 

\acknowledgments

D. Dantchev acknowledges the  hospitality of Max-Planck-Institute for Metals Research in
Stuttgart  as well as the financial support of the Alexander von Humboldt
Foundation.

\appendix
\section{The correlation length amplitude relation}
\label{clar} The coordinate transformation given by
Eq.(\ref{coorlam}) removes the anisotropy from the spin model
defined by Eq.(\ref{Hamlam}) in the vicinity of the critical
point. From the coordinate transformation and the principle of two
scale - factor universality a relation between the correlation
length amplitudes $\xi_{\parallel,0}(\lambda)$ and
$\xi_{\perp,0}(\lambda)$ (see Eq.(\ref{xilam})) can be
established, which will be derived in the following under the
assumption that hyperscaling is also valid, e.g., for $2 < d < 4$
and short-ranged interactions.

According to the coordinate transformation given by
Eq.(\ref{coorlam}) we obtain $\xi'_\parallel = \xi_\parallel$ and
$\xi'_\perp = R(\lambda) \ \xi_\perp = \xi_\parallel$ (see
Eq.(\ref{Llam})), i.e., the parallel correlation length
$\xi_\parallel$ of the (untransformed) anisotropic system remains
as the only correlation length of the (transformed) isotropic
system. According to the principles of two scale - factor
universality and hyperscaling the singular part of the (bulk) free
energy density $f'_{b,sing}(t)$ of the transformed spin system can
then be written in the form
\begin{equation} \label{fbsingprime}
f'_{b,sing}(\lambda,t) = {\cal A} \left[\xi_\parallel(\lambda,t)\right]^{-d},
\end{equation}
where $t = \beta_c(\lambda)/\beta - 1$ is the reduced temperature and
${\cal A}$ is a {\em universal}
amplitude. Strictly speaking, one has to distinguish between {\em two}
universal amplitudes ${\cal A}_+$ for $T > T_c$ and ${\cal A}_-$ for
$T < T_c$. We disregard this distinction in Eq.(\ref{fbsingprime}) in
order to simplify the notation. Eq.(\ref{fbsingprime}) is valid for
$T > T_c$ and $T < T_c$ {\em separately}, provided, the correlation length
remains finite for $T < T_c$. According to Eq.(\ref{coorlam}) the unit
volume $v$ of the system transforms as
\begin{equation} \label{vprime}
v' = R(\lambda) v
\end{equation}
and therefore we find
\begin{eqnarray} \label{fbsing}
f_{b,sing}(\lambda,t) &=& R(\lambda)\  f'_{b,sing}(\lambda,t)
\nonumber\\
&=& {\cal A} R(\lambda)\ \left[\xi_\parallel(\lambda,t)\right]^{-d}
\nonumber\\
&=& {\cal A} R(\lambda)\ \left[\xi'_\perp(\lambda,t)\right]^{-1}
\left[\xi'_\parallel(\lambda,t)\right]^{-(d-1)}
\nonumber\\
&=& {\cal A} \left[\xi_\perp(\lambda,t)\right]^{-1}
\left[\xi_\parallel(\lambda,t)\right]^{-(d-1)}
\end{eqnarray}
for the singular part of the bulk free energy density of the anisotropic,
i.e., the untransformed system.

According to Eq.(\ref{xilam}) we have the alternative form
\begin{eqnarray} \label{fbsingalt}
f_{b,sing}(\lambda,t) &=& {\cal A} \left[\xi_{\perp,0}(\lambda)\right]^{-1}
\left[\xi_{\parallel,0}(\lambda)\right]^{-(d-1)}\ |t|^{d\nu}
\nonumber \\
&\equiv& A(\lambda)\ |\beta_c(\lambda)/\beta - 1|^{d\nu}
\end{eqnarray}
for Eq.(\ref{fbsing}). The nonuniversal amplitude $A(\lambda)$ and
$\beta_c(\lambda)$ must be independent of the labelling of the
lattice axes, i.e., the direction which is chosen to be the
'perpendicular' one. From this symmetry argument and the
particular choice of the coupling constants $J_\parallel(\lambda)$
and $J_\perp(\lambda)$ in Eq.(\ref{Jlam}) we have already obtained
$\beta'(\lambda = 0) = 0$ in Eq.(\ref{Tc0}). Likewise, we obtain
$A'(\lambda = 0) = 0$ from this symmetry argument. We therefore
conclude that
\begin{equation} \label{fbsinglam0}
\left. \frac{d}{d\lambda} f_{b,sing}(\lambda,t) \right|_{\lambda = 0} = 0
\end{equation}
and that due to
\begin{equation} \label{Alam}
A(\lambda) = {\cal A} \left[\xi_{\perp,0}(\lambda)\right]^{-1}
\left[\xi_{\parallel,0}(\lambda)\right]^{-(d-1)}
\end{equation}
one also concludes from $A'(\lambda = 0) = 0$ that
\begin{equation} \label{ddlamxi}
\left. \frac{d}{d\lambda} \left\{
\left[\xi_{\perp,0}(\lambda)\right]^{-1}
\left[\xi_{\parallel,0}(\lambda)\right]^{-(d-1)} \right\}
\right|_{\lambda = 0} = 0.
\end{equation}
From Eq.(\ref{ddlamxi}) we finally obtain the important correlation length
amplitude relation
\begin{equation} \label{xi0relat}
(d - 1)
\left.\frac{d}{d\lambda}\ \xi_{\parallel,0}(\lambda)\right|_{\lambda=0}
+ \left.\frac{d}{d\lambda}\ \xi_{\perp,0}(\lambda)\right|_{\lambda=0} = 0
\end{equation}
which is needed in the derivation of the stress tensor
representation of the Casimir force for lattice spin models
presented in Appendix \ref{ast}.

\section{The stress tensor representation of the Casimir force}
\label{ast} The derivative of the excess free energy $f_{\rm ex}$
with respect to $\lambda$ at the isotropic point $\lambda = 0$ is
given by Eq.(\ref{fexlam}) in the main text. The relation between
Eq.(\ref{fexlam}) and the Casimir force defined by Eq.(\ref{def})
yields a lattice expression of the stress tensor. This will be
investigated here in the critical regime. Above the critical
temperature all expressions will be exponentially small and can be
neglected. Below the critical temperature Goldstone modes in $O(N
\geq 2)$ systems also give rise to algebraically decaying
finite-size effects, which will not be considered here.

In order to find the relation between Eqs.(\ref{def}) and (\ref{fexlam})
we use the coordinate transformation given by Eq.(\ref{coorlam}) and note
that unlike the unit volume $v$ (see Eq.(\ref{vprime})) the unit area remains
{\em invariant} under Eq.(\ref{coorlam}). We recall that in the transformed
(isotropic) system we have $\xi'_{\perp,0}(\lambda) =
\xi'_{\parallel,0}(\lambda) = \xi_{\parallel,0}(\lambda)$ and we therefore
find
\begin{eqnarray} \label{fexLperp}
\left.\frac{d f_{\rm ex}}{d\lambda} \right|_{\lambda=0} &=&\nonumber \\
\left.\frac{d f'_{\rm ex}}{d\lambda} \right|_{\lambda=0} &=&
\left.\frac{\partial f'_{\rm ex}}{\partial L_\perp'}
\right|_{\lambda=0} \left.\frac{d L_\perp'}{d\lambda}
\right|_{\lambda=0} + \left.\frac{\partial f'_{\rm ex}}{\partial
\xi'_{\perp,0}} \right|_{\lambda=0}
\left.\frac{d \xi'_{\perp,0}}{d\lambda} \right|_{\lambda=0} \nonumber \\
&=& - \beta F_{Casimir}\ R'(0)\ L_\perp\\
&&+ \left.\frac{\partial f'_{\rm ex}}{\partial
\xi'_{\perp,0}} \right|_{\lambda=0}
\left.\frac{d \xi'_{\perp,0}}{d\lambda} \right|_{\lambda=0}
\nonumber
\end{eqnarray}
where Eqs.(\ref{def}) and (\ref{Llam}) have been used.
In order to evaluate the derivative $\partial f'_{\rm ex} /
\partial \xi'_{\perp,0}$ in the critical regime, we use the critical
finite-size scaling form
\begin{equation}\label{fexsingprime}
f'_{\rm ex}(t,L'_\perp) = \left. L'\right._\perp^{-(d-1)} g'_{\rm
ex}\left[\ t\ \left(L'_\perp / \xi'_{\perp,0}\right)^{1/\nu}
\right]
\end{equation}
and disregard the exponentially small contributions to
Eq.(\ref{fexLperp}) from the regular part of the excess free
energy. Note, that for periodic boundary conditions the free
energy of the finite system $f'(t,L'_\perp)$ can be decomposed, as
usual, in a regular $f'_{\rm reg}(t,L'_\perp)$ and a singular
$f'_{\rm sing}(t,L'_\perp)$ parts, where the regular part $f'_{\rm
reg}(t,L'_\perp)$ can be taken to be equal (up to, eventually,
exponentially small corrections) to that one of the infinite
system, i.e. $f'_{\rm reg}(t,L'_\perp)=f'_{\rm reg}(t,\infty)$
\cite{privman90}. That is why, for the periodic boundary
conditions, the above equation (\ref{fexsingprime}) is valid for
the {\it total} excess free energy (and not only for its singular
part). From Eq.(\ref{fexsingprime}), we immediately obtain
\begin{equation}\label{fexsing0}
\left. \frac{\partial f'_{\rm ex}}{\partial \xi'_{\perp,0}}
\right|_{\lambda = 0} = - \frac{t}{\nu\ \xi_0} \frac{\partial
f'_{\rm ex}}{\partial t},
\end{equation}
where all terms on the r.h.s. of Eq.(\ref{fexsing0}) have already
been evaluated at $\lambda = 0$. To further evaluate
Eq.(\ref{fexsing0}) we note that the excess internal energy
$u_{\rm ex}$   is given by
\begin{equation}\label{uex}
u_{\rm ex} = \frac{\partial f_{\rm ex}}{\partial \beta} =
-\frac{\beta_c(0)}{\beta^2}\frac{\partial f_{\rm ex}}{\partial t}.
\end{equation}
In the vicinity of $\beta = \beta_c(0)$ Eq.(\ref{fexsing0}) can be rewritten
as
\begin{equation}\label{fexsing0uex}
\left. \frac{\partial f'_{\rm ex}}{\partial \xi'_{\perp,0}}
\right|_{\lambda = 0} = \frac{1}{\nu\ \xi_0} (\beta_c(0) - \beta)\
u_{\rm ex}.
\end{equation}
In order to evaluate the derivative
$d \xi'_{\perp,0} / d\lambda |_{\lambda=0}$
in Eq.(\ref{fexLperp}) we note that according to
Eq.(\ref{coorlam}) we have $\xi'_{\perp,0}(\lambda) = \xi_{\parallel,0}
(\lambda)$. From the definition of $R(\lambda)$ given by Eq.(\ref{Llam})
we find by taking the derivative $R'(\lambda)$ with respect to $\lambda$
at $\lambda = 0$
\begin{equation} \label{Rprime0}
R'(0) = \frac{1}{\xi_0} \left[
\left. \frac{d \xi_{\parallel,0}}{d \lambda} \right|_{\lambda=0}
- \left. \frac{d \xi_{\perp,0}}{d \lambda} \right|_{\lambda=0}
\right].
\end{equation}
We eliminate $d\xi_{\perp,0}/d\lambda |_{\lambda=0}$ from
Eq.(\ref{Rprime0}) using Eq.(\ref{xi0relat}) of Appendix
\ref{clar} and obtain
\begin{equation}\label{xipar0}
\left. \frac{d \xi_{\parallel,0}}{d \lambda} \right|_{\lambda=0}
= \frac{\xi_0}{d}\ R'(0).
\end{equation}
We finally insert Eqs.(\ref{fexsing0uex}) and (\ref{xipar0}) into
Eq.(\ref{fexLperp}) and, by rearranging the terms, we  obtain for
the Casimir force
\begin{eqnarray} \label{FCasimir}
\beta F_{Casimir} &=& \left[ R'(0) \right]^{-1} \lim_{L_\parallel \to \infty}
\frac{\beta J'(1)}{L_\parallel^{d-1} L_\perp}
\nonumber \\
&\times& \left\langle \sum_{\bf R} \left[\sum_{k=1}^{d-1} S_{\bf
R} S_{{\bf R}+{\bf e}_k} - (d-1) S_{\bf R} S_{{\bf R}+{\bf e}_d}
\right]\right\rangle, \nonumber \\
&&  +\frac{1}{d\nu}\ (\beta_c(0)-\beta)\ \frac{u_{\rm ex}}{L_\perp}
\nonumber \\
&\equiv& \langle t_{\perp \perp} \rangle
\end{eqnarray}
where Eq.(\ref{fexlam}) has also been used.
Note that the first term in Eq.(\ref{FCasimir}) is generated by the
anisotropy variation whereas the second term originates from a change
in length scales enforced by the coordinate transformation given by
Eq.(\ref{coorlam}),

In order to express
Eq.(\ref{FCasimir}) as the thermal average $\langle t_{\perp \perp} \rangle$
of the normal component of the stress tensor we note that
\begin{equation} \label{uub}
u_{\rm ex} / L_\perp = u - u_b\ ,
\end{equation}
where $u$ is the volume energy density of the slab and $u_b$ the
volume energy density in the bulk. Naturally, $u$ and $u_b$ are
thermal averages of properly normalized Hamiltonians ${\cal
\hat{H}}$ and ${\cal \hat{H}}_b$. More specifically, ${\cal
\hat{H}}$ is the Hamiltonian of the finite system normalized per
unit volume, while ${\cal \hat{H}}_b$ is the corresponding
Hamiltonian for the bulk system (i.e. one imagines an arbitrary
finite connected region of spins whose mutual probability
distribution is obtained by taking the thermodynamic limit while
integrating out all spins {\em outside} that fixed region. This is
done for any finite region of the lattice). From
Eqs.(\ref{FCasimir}) and (\ref{uub}) the operator form of the
stress tensor given by Eq.(\ref{st}) in the main text can then be
read off.

\section{The Two-Dimensional Ising Model}
\label{aIsing}

As it has been shown in the main text, see Eq. (\ref{stcrIsing}),
in the critical region of the finite system the stress tensor is
given by
\begin{eqnarray}\label{astcrIsing}
t_{x,x}(i,j)&=&\frac{1}{2\sqrt{2}}(S_{i,j}S_{i,j+1}-S_{i,j}S_{i+1,j})\nonumber \\
&&+ \frac{1}{2}(\beta_c-\beta)({\cal \hat{H}}-{\cal \hat{H}}_b).
\end{eqnarray}

Let us now calculate the variance of the stress tensor $\Delta
t_{x,x}(i,j)$, which we will interpret as a variance of a local
measurement of the Casimir force made near the point $(i,j)$. For
the leading behavior of the variance near $T_c$ one has
\begin{eqnarray}
\Delta t_{x,x}(i,j) &=& \frac{1}{2} \langle
S_{i,j}^2(S_{i,j+1}-S_{i+1,j})^2 \rangle -\langle t_{x,x}(i,j)
\rangle^2 \nonumber \\
 & = & 1-\langle S_{i,j+1} S_{i+1,j} \rangle - \langle t_{x,x}(i,j)
\rangle^2. \label{avarIsing}
\end{eqnarray}
Obviously, it holds that $\langle S_{i,j+1} S_{i+1,j}
\rangle=\langle S_{0,1}S_{1,0}\rangle=\langle S_{0,0} S_{1,1}
\rangle$, because of the symmetry of the Ising model on a square
lattice under periodic boundary conditions. The correlations
$\langle S_{0,0} S_{1,1} \rangle$ are well known for the bulk
system \cite{CW73}:
\\ i) for $T<T_c$
\begin{equation}
\langle S_{0,0} S_{1,1} \rangle=\frac{2}{\pi}{\bf
E}\left(\frac{1}{u}\right).
\end{equation}
ii) for $T>T_c$
\begin{equation}
\langle S_{0,0} S_{1,1} \rangle=\frac{2}{\pi u}\left[ {\bf
E}(u)+(u^2-1){\bf K}(u) \right].
\end{equation}
where, according to Eq. (\ref{Cri2dI}),
\begin{equation}
u=\sinh(2\beta J_x)\sinh(2\beta J_y).
\end{equation}
iii) for $T=T_c$, which is given by $u=1$, it follows that
$\langle S_{0,0} S_{1,1} \rangle=2/\pi$.

In the above expressions $\bf K$ and $\bf E$ are the complete
elliptic integrals of first, and of second kind, respectively.
From them and Eq. (\ref{avarIsing}) one easily obtains expressions
for the behavior of the variance $\Delta t_{x,x}(i,j)$ of the
stress tensor below, above, and at $T_c$. At $T=T_c$, for example,
one has that
\begin{equation}
\label{avarsingleIsing}
\Delta t_{x,x}(i,j)\simeq 1-2/\pi.
\end{equation}
Definitely, in addition from the above nonuniversal part the
variance contains also an universal parts that  are negligible in
comparison with the nonuniversal one.

An estimation can be also derived for
$\Delta\sum_{i,j}t_{x,x}(i,j)$. With a variance of such a type one
deals when, say, Monte Carlo simulations of the force are
performed. According to Eqs. (\ref{htdef}) and (\ref{tt}), at
$T=T_c$
\begin{eqnarray}
\label{aIvst}
    \Delta\sum_{i,j}\tilde{t}_{x,x}(i,j)&=& \frac{\partial^2}{\partial \lambda^2}\left[\ln \sum
e^{-\beta {\cal H}(\lambda)} \right]
\\
\nonumber  &=& M L \frac{\partial^2}{\partial \lambda^2}[-\beta
\tilde{f}(T_c,\lambda)],
\end{eqnarray}
and, therefore, from Eq. (\ref{stcrIsing}) it follows that
\begin{equation}\label{avstIsing}
\Delta\sum_{i,j}t_{x,x}(i,j)=\frac{1}{2\ln^2(1+\sqrt{2})} M L
\frac{\partial^2}{\partial \lambda^2}[-\beta
\tilde{f}(T_c,\lambda)].
\end{equation}
It is clear that the leading order behavior of the variance will
stem from the bulk contribution to the free energy - the
finite-size terms will produce only corrections to it. The bulk
free energy of the anisotropic two-dimensional Ising model is well
known (see, e.g., \cite{CW73})
\begin{eqnarray}
  -\beta f &=& \ln 2+\frac{1}{2}\int_{-\pi}^{\pi}\frac{d\theta_1}{2\pi} \int_{-\pi}^{\pi}\frac{d\theta_2}{2\pi}
  \ln\left[\cosh (2\beta J_x) \right. \nonumber \\
   &&\left. \times \cosh(2\beta J_y)-\sinh(2\beta
   J_x)\cos(\theta_1)\right.\nonumber \\
   && \left. - \sinh(2\beta J_y)\cos(\theta_2) \right].
\end{eqnarray}
Setting here $J_x=(1+\lambda)J$ and $J_x=(1-\lambda)J$, we
immediately obtain $-\beta \tilde{f}(\beta J,\lambda)$, and from
(\ref{avstIsing}) one then derives (at $T=T_c$) that
\begin{eqnarray}\label{avstIsingfinal}
\lefteqn{\Delta\sum_{i,j}t_{x,x}(i,j)=}\\
&&
\frac{1}{2}\left[-\frac{1}{2}+\frac{1}{\pi^2}\int_{-\pi}^{\pi}d\theta_1
\int_{-\pi}^{\pi}d\theta_2
\frac{\sin^2(\frac{\theta_1}{2})\sin^2(\frac{\theta_2}{2})}
{\left(1-\frac{\cos(\theta_1)+\cos(\theta_2)}{2}\right)^2}\right]
\nonumber\\
&=&\frac{1}{2}\left[-\frac{1}{2}+\frac{2}{\pi}\right] M L \simeq
0.068 M L. \nonumber
\end{eqnarray}
We again observe that the variance of the sum of $t_{x,x}(i,j)$ is
proportional to the total number of summands  in this sum. This is
the result given in Eq. (\ref{vstIsingfinal}) in the main  text.

\section{The Spherical Model}
\label{aspm}

Using the identity
\begin{equation}
\ln(1+z)=\int_{0}^{\infty}\frac{dx}{x}\left(1-e^{- z
x}\right)e^{-x},
\end{equation}
the equation for the free energy (\ref{freeenergyspmdef}) becomes
\begin{equation}\label{freeenergy}
\beta f(K,{\bf
N})=\frac{1}{2}\left[\ln{\frac{K}{2\pi}}-K\right]+\sup_{w>0}\left\{U(w,{\bf
N})-\frac{1}{2}Kw\right\},
\end{equation}
where
\begin{eqnarray}\label{u}
    \lefteqn{U(w,{\bf N})=\frac{1}{2N} \sum_{{\bf k}\in {\cal B}_\Lambda}
\ln{\left[ w+1-\frac{\hat{J}({\bf k})}{\hat{J}({\bf 0
})}\right]}} \\
&=& \frac{1}{2}\int_0^\infty \frac{dx}{x}\left(e^{-x}-e^{-x
w}\frac{1}{N}\sum_{{\bf k}\in {\cal
B}_\Lambda}e^{-x[1-\hat{J}({\bf k})/\hat{J}({\bf 0 })]}
\right).\nonumber
\end{eqnarray}
The supremum is attained at the value of $w$ that is a solution of
the (spherical field) equation
\begin{equation}\label{eq}
\frac{1}{N}\sum_{{\bf k}\in {\cal B}_\Lambda}\int_0^\infty e^{-x
w} e^{-x[1-\hat{J}({\bf k})/\hat{J}({\bf 0 })]} dx=K.
\end{equation}

 For nearest neighbor interactions the Fourier
transform of the interaction reads
\begin{equation}
\hat{J}({\bf k})=2\sum_{j=1}^d J_j\cos(k_j a_j).
\end{equation}
Then, for the spherical field equation and the sum $U(w,{\bf N})$,
we obtain
\begin{equation}\label{eqsr}
    \int_0^\infty e^{- x w}\left[\prod_{j=1}^d \frac{1}{N_j}\sum_{k_j}e^{-x b_j (1-\cos k_j
    a_j)}\right]=K,
\end{equation}
\begin{eqnarray}\label{usr}
\lefteqn{U(w,{\bf N})= }\\
&& \frac{1}{2}\int_0^\infty \frac{dx}{x}\left(e^{-x}-e^{-x
w}\left[\prod_{j=1}^d \frac{1}{N_j}\sum_{k_j}e^{-x b_j (1-\cos k_j
    a_j)}\right]\right),\nonumber
\end{eqnarray}
where $b_j=J_j/\sum_{j=1}^d J_j$. Using the identity \cite{SP85}
\begin{equation}\label{identitySP}
    \sum_{n=0}^{N-1}\exp\left[x\cos\frac{2\pi
    n}{N}\right]=N\sum_{q=-\infty}^{\infty}I_{q N}(x),
\end{equation}
Eqs. (\ref{eqsr}) and (\ref{usr}) can be written in the form
\begin{equation}\label{eqsrI}
\int_{0}^\infty e^{-x w}    \prod_{j=1}^{d}\left[e^{-x
b_j}\sum_{q_j=-\infty}^\infty I_{q_jN_j}(b_j x)
    \right]dx=K,
\end{equation}
and
\begin{eqnarray}\label{usrI}
\lefteqn{U(w,{\bf N})= }\\
&& \frac{1}{2}\int_0^\infty \frac{dx}{x}\left(e^{-x}-e^{-x
w}\prod_{j=1}^{d}\left[e^{-x b_j}\sum_{q_j=-\infty}^\infty
I_{q_jN_j}(b_j x)
    \right]\right).\nonumber
\end{eqnarray}

In analogical way one can consider the behavior of the bulk
system. Then, in the limit $N_j\rightarrow \infty$,
$j=1,\cdots,d$, one obtains the bulk equation for the spherical
field
\begin{eqnarray}\label{eqsrIb}
K &=& \int_{0}^\infty e^{-x w}    \prod_{j=1}^{d}\left[e^{-x b_j}
I_{0}(b_j x)
    \right]dx \\
    &=& \frac{1}{(2\pi)^d}\int_0^{2\pi}dn_1\cdots
    \int_0^{2\pi}dn_d \frac{1}{w+\sum_{j=1}^d b_j(1-\cos n_j)},\nonumber
\end{eqnarray}
and the following contribution into the free energy
\begin{eqnarray}\label{ub}
U_b(w) &=& \frac{1}{2}\int_0^\infty \frac{dx}{x}\left(e^{-x}-e^{-x
w}\prod_{j=1}^{d}\left[e^{-x b_j} I_{0}(b_j x)
    \right]\right) \nonumber \\
    &=& \frac{1}{2}\frac{1}{(2\pi)^d}\int_0^{2\pi}dn_1\cdots
    \int_0^{2\pi}dn_d \nonumber \\
    && \times \ln\left[w+\sum_{j=1}^d b_j(1-\cos
    n_j)\right].
\end{eqnarray}

In a similar way, starting from Eq. (\ref{cordef}), one can show
that the bulk two-point correlation function in such an
anisotropic system is
\begin{eqnarray}\label{cor}
G({\bf r},t)&=&\frac{1}{w\beta \hat{J}({\bf 0})}\int_0^\infty
d\rho \ e^{-\rho} \prod_{j=1}^d \exp\left(\rho \frac{b_j}
{w}\right) \nonumber \\
& & \times \frac{1}{2\pi} \int_0^{2\pi} \exp{\left[ i n_j l_j
+\rho \frac{b_j}{w}\cos n_j\right]}.\ \
\end{eqnarray}
Supposing that $l_j \gg 1$, $j=1,\cdots,d$, from (\ref{cor}) one
obtains
\begin{equation}\label{corr}
G({\bf r}, t)\simeq \frac{1}{w\beta \hat{J}({\bf 0})}\int_0^\infty
d\rho \ e^{-\rho} \prod_{j=1}^d \frac{\exp\left[-\frac{w}{2\rho
b_j}l_j^2\right]}{\sqrt{\rho b_j/w}},
\end{equation}
wherefrom one concludes that the correlation length $\xi_j$ in
direction $j$ is
\begin{equation}
\label{corrlength} \xi_j=\sqrt{2b_j/w}
\end{equation}
with the critical point of the system given by $w=0$ (note that in
the spherical model, because of the so-called equation of the
spherical field, Eq. (\ref{eqsrIb}), $w$ depends on the coupling
$K$, dimensionality $d$ and on the anisotropy described by the
constants $b_j$, $j=1, \cdots, d$). Therefore, one has
\begin{equation}\label{acorrl}
    \frac{\xi_j}{\xi_i}=\sqrt{\frac{b_j}{b_i}}=\sqrt{\frac{J_j}{J_i}}.
\end{equation}
Taking $J_j$, $j=1,\cdots,j$ in the form prescribed by Eq.
(\ref{Jlam}) one obtains
\begin{equation}\label{der}
    \frac{d}{d\lambda}\left.\left(\frac{\xi_{\parallel,0}}{\xi_{\perp,0}}\right)\right
    |_{\lambda=0}=\frac{d}{2} \frac{J'(1)}{J(1)},
\end{equation}
and, thus, making use of Eq. (\ref{st}), we derive the explicit
form of the stress tensor within the spherical model
\begin{eqnarray}
\label{astsm}
t_{\perp \perp}({\bf R}) &=& \frac{\beta J}{d/2}
\left[\sum_{k=1}^{d-1} S_{\bf R} S_{{\bf R}+{\bf e}_k} - (d-1)
S_{\bf R} S_{{\bf R}+{\bf e}_d}\right]\nonumber \\
&& +\frac{1}{d \; \nu}(\beta_c-\beta)\left({\cal \hat{H}}- {\cal
\hat{H}}_b\right),
\end{eqnarray}
where ${\cal \hat{H}}$ is the Hamiltonian (normalized per unit
particle) of the finite, and ${\cal \hat{H}}_b$ of the infinite
system.

\subsubsection{Evaluation of the finite-size excess free energy of the anisotropic system}
From Eqs. (\ref{usrI}) and (\ref{ub}) one has
\begin{equation}\label{udec}
    U(w,{\bf N})=U_b(w)+\Delta U(w,{\bf N}),
\end{equation}
where
\begin{equation}\label{du}
\Delta U(w,{\bf N})=-\frac{1}{2}\sum_{{\bf q}\ne {\bf 0}}
\int_0^{\infty}\frac{dx}{x}e^{-x w}\prod_{j=1}^d e^{-x b_j}I_{q_j
N_j}(x b_j).
\end{equation}
Next, with the help of the expansion \cite{SP85}
\begin{equation}\label{SP85}
    I_\nu(x)=\frac{\exp(x-\nu^2/2x)}{\sqrt{2\pi x}} (1+\frac{1}{8 x}+\frac{9-32\nu^2}{2! (8x)^2}+ \ \cdots)
\end{equation}
$\Delta U(w,{\bf N})$ can be cast in the form
\begin{equation}\label{deltau}
    \Delta U(w,{\bf N})=-\frac{1}{2}\sum_{{\bf q}\ne {\bf
    0}}\int_0^\infty \frac{dx}{x}e^{-x w}\prod_{j=1}^d \frac{e^{-N_j^2 q_j^2/2xb_j}}{\sqrt{2\pi
    xb_j}},
\end{equation}
wherefrom, in the limit of a film geometry $N_1, N_2, \cdots
N_{d-1} \rightarrow \infty$, with $N_d=N_\perp$, one obtains
\begin{eqnarray}\label{deltauscaling}
\Delta
U(w,N_\perp)&=&-\frac{2}{(2\pi)^{d/2}}\left(\frac{b_\perp}{b_\parallel}\right)^{(d-1)/2}y^{d/4}\nonumber
\\ && \times
\sum_{q=1}^{\infty}\frac{K_{d/2}(q\sqrt{y})}{q^{d/2}}N_\perp^{-d}.
\end{eqnarray}
Here we have taken $J_1=J_2=\cdots J_{d-1}=J_\parallel$ and
$J_d=J_\perp$, which corresponds to $b_1=b_2=\cdots
b_{d-1}=b_\parallel$ and $b_d=b_\perp$, whereas
$y=2wN_\perp^2/b_\perp$.

All what remains now is to deal with the behavior of the bulk term
$U_b(w)$ when $K$ is close to $K_c$, i.e. when $w<<1$. This
analysis is well known for the isotropic case, here we will, very
briefly, extend it to cover the anisotropic case also. Starting
from Eq. (\ref{ub}), one obtains
\begin{equation}\label{ubdec}
    U_b(w)=U_b(0)+\frac{1}{2}\int_0^w d \omega W_d(\omega|{\bf
    b}),
\end{equation}
where
\begin{equation}\label{ub0}
    U_b(0)=\frac{1}{2}\int_0^\infty \frac{dx}{x}\left[e^{-x}-\prod_{j=1}^d e^{-xb_j}I_0(xb_j) \right]
\end{equation}
is a temperature independent constant and
\begin{equation}\label{wb}
    W_d(\omega|{\bf b})=\int_0^\infty dx e^{-x \omega} \prod_{j=1}^d
    e^{-x b_j} I_0(x b_j)
\end{equation}
is the generalized Watson-type integral (the standard one is with
$b_j=b$ for all $j=1,\cdots,d$). Using the standard technique for
evaluation of such type of integrals (see, e.g., \cite{BDT00}) one
derives that, for $2<d<4$,
\begin{equation}\label{wbfinal}
    W_d(\omega|{\bf b})\simeq W_d(0|{\bf
    b}),
\end{equation}
wherefrom it follows that, again for $2<d<4$,
\begin{equation}\label{ubfinal}
    U_b(w) \simeq U_b(0)+\frac{1}{2}w W_d(0|{\bf
    b})-\frac{1}{2}\frac{\Gamma(-d/2)}{(2\pi)^{d/2}\prod_{j=1}^{d}\sqrt{b_j}}
    \omega^{d/2}.
\end{equation}
Taking into account that in terms of the "anisotropic" Watson
integral the equation of the spherical field simply is
\begin{equation}\label{sfeq}
    K=W_d(w|{\bf b}),
\end{equation}
and that, according to Eq. (\ref{corrlength}), the critical point
is fixed by $w=0$, we conclude that the critical coupling  of the
anisotropic system is
\begin{equation}\label{acritT}
    K_c \equiv 2\beta_c \sum_{j=1}^d J_j=W_d(0|{\bf b})=
    \int_0^\infty dx \prod_{j=1}^d e^{-x b_j}I_0(x b_j).
\end{equation}
Then, close to $K=K_c$, for the free energy density of bulk system
from Eqs. (\ref{freeenergy}), (\ref{ubfinal}) and (\ref{critT})
one obtains
\begin{eqnarray}
\label{bulkfreeenergy}
  \beta f_b(K|{\bf b}) &=& \frac{1}{2}\left[\ln \frac{K}{2\pi}-K\right]
  +\frac{1}{2}w_b(K_c-K)+U_b(0)\nonumber \\
   &&-\frac{1}{2}\frac{\Gamma(-d/2)}{(2\pi)^{d/2}\prod_{j=1}^{d}\sqrt{b_j}}
    \omega^{d/2},
\end{eqnarray}
where, for $K\le K_c$, the parameter $w_b$ is the solution of the
equation
\begin{equation}\label{sfebulk}
K=K_c
+\frac{\Gamma(1-d/2)}{(2\pi)^{d/2}\prod_{j=1}^{d}\sqrt{b_j}}\
    \omega_b^{d/2-1}
\end{equation}
whereas for $K>K_c$ the supremum of the free energy is attained at
$w_b=0$. Similarly, for the free energy density of the {\it
finite} system from Eqs. (\ref{freeenergy}), (\ref{deltau}),
(\ref{ubfinal}) and (\ref{critT}) we obtain
\begin{eqnarray}
\label{finitesizefreeenergy}
  \beta f(K,{\bf N}|{\bf b}) &=& \frac{1}{2}\left[\ln \frac{K}{2\pi}-K\right]
  +\frac{1}{2}w(K_c-K)+U_b(0)\nonumber \\
   &&-\frac{1}{2}\frac{\Gamma(-d/2)}{(2\pi)^{d/2}\prod_{j=1}^{d}\sqrt{b_j}}
    \omega^{d/2}\\
    && -\frac{1}{2}\sum_{{\bf q}\ne {\bf
    0}}\int_0^\infty \frac{dx}{x}e^{-x w}\prod_{j=1}^d \frac{e^{-N_j^2 q_j^2/2xb_j}}{\sqrt{2\pi
    xb_j}},\nonumber
\end{eqnarray}
where $w$ is the solution of the finite-size equation for the
spherical field
\begin{eqnarray}\label{sfefinite}
K &=& K_c
+\frac{\Gamma(1-d/2)}{(2\pi)^{d/2}\prod_{j=1}^{d}\sqrt{b_j}}\
    \omega^{d/2-1}\nonumber \\
&& +
    \sum_{{\bf q}\ne {\bf
    0}}\int_0^\infty dx e^{-x w}\prod_{j=1}^d \frac{e^{-N_j^2 q_j^2/2xb_j}}{\sqrt{2\pi
    xb_j}}.
\end{eqnarray}
Recalling that, for $2<d<4$, the spherical model has a critical
exponent $\nu=1/(d-2)$ one can, in the limit of a film geometry
$N_1, N_2, \cdots, N_{d-1}\rightarrow \infty$, from Eqs.
(\ref{deltauscaling}), (\ref{bulkfreeenergy}) and
(\ref{finitesizefreeenergy}), obtain an expression for the {\it
excess} free energy $\beta(f-f_b)$ (per spin)
\begin{eqnarray}\label{aexcess}
\lefteqn{\beta \left[f(K, N_\perp|{\bf  b})-f_b(K|{\bf b})\right]=
\left\{\frac{1}{4}x_1 (y-y_\infty)\right.}
\\
&& \left. -\frac{1}{2}\frac{\Gamma(-d/2)}{(4\pi)^{d/2}}
\left(\frac{b_\perp}{b_\parallel}\right)^{(d-1)/2}\left(y^{d/2}-y_\infty^{d/2}\right)\right.\nonumber
\\
&&\left. -\frac{2}{(2\pi)^{d/2}}
\left(\frac{b_\perp}{b_\parallel}\right)^{(d-1)/2}
y^{d/4}\sum_{q=1}^{\infty}\frac{K_{d/2}(q\sqrt{y})}{q^{d/2}}\right\}N_\perp^{-d}
\nonumber
\end{eqnarray}
in a scaling form. In the above equation 
\begin{equation}
\label{ax1} x_1=b_\perp (K_c-K)N_\perp^{1/\nu}, \
\nu=\frac{1}{d-2},
\end{equation}
 is the
temperature scaling variable, $y_\infty=2w_b N_\perp^2/b_\perp$ is
the solution of the bulk spherical field equation
\begin{equation}\label{asfsfeb}
    -\frac{1}{2}x_1=\frac{\Gamma(1-d/2)}{(4\pi)^{d/2}}\left(\frac{b_\perp}{b_\parallel}\right)^{(d-1)/2}
    y_\infty^{d/2-1},
\end{equation}
while $y=2wN_\perp^2/b_\perp$ is the solution of the finite-size
spherical field equation
\begin{eqnarray}\label{asfsfefinite}
   \lefteqn{ -\frac{1}{2}x_1 = \frac{\Gamma(1-d/2)}{(4\pi)^{d/2}}\left(\frac{b_\perp}{b_\parallel}\right)^{(d-1)/2}
    y^{d/2-1}}\\
    && + \frac{2}{(2\pi)^{d/2}}
\left(\frac{b_\perp}{b_\parallel}\right)^{(d-1)/2}
y^{d/4-1/2}\sum_{q=1}^{\infty}\frac{K_{d/2-1}(q\sqrt{y})}{q^{d/2-1}}.
\nonumber
\end{eqnarray}
For the Casimir force (see also Eq. (\ref{def}))
\begin{equation}\label{casforce}
\beta F_{{\rm Casimir}}=-\frac{\partial}{\partial
N_\perp}\left[N_\perp \beta (f-f_b) \right]
\end{equation}
from Eqs. (\ref{excess}), (\ref{sfsfeb}), (\ref{sfsfefinite}) one
obtains
\begin{eqnarray}
\label{acassm}
 \lefteqn{\beta F_{{\rm Casimir}} = N_\perp^{-d}\left\{ \frac{1}{4}x_1 (y-y_\infty)\right.} \\
 && \left.-(d-1)\left(\frac{b_\perp}{b_\parallel}\right)^{(d-1)/2}\left[\frac{1}{2}\frac{\Gamma(-d/2)}{(4\pi)^{d/2}}
\left(y^{d/2}-y_\infty^{d/2}\right)\right. \right. \nonumber \\
  &&\left. \left. + \frac{2}{(2\pi)^{d/2}}
y^{d/4}\sum_{q=1}^{\infty}\frac{K_{d/2}(q\sqrt{y})}{q^{d/2}}\right]
\right\}. \nonumber
\end{eqnarray}
We are ready now to determine the Casmir amplitudes $\Delta$ in
the spherical model. Having in mind Eq. (\ref{cadef}), at $K=K_c$
(then $y_\infty=0$), for the isotropic system (then
$b_\perp=b_\parallel=1/d$) one obtains from Eq. (\ref{cassm})
\begin{equation}\label{casamplsm}
    \Delta=-\left[\frac{1}{2}\frac{\Gamma(-d/2)}{(4\pi)^{d/2}} y_c^{d/2}+\frac{2}{(2\pi)^{d/2}}
 y_c^{d/4}\sum_{q=1}^{\infty}\frac{K_{d/2}(q\sqrt{y_c})}{q^{d/2}}\right],
\end{equation}
where
\begin{equation}\label{criti}
\frac{\Gamma(-d/2)}{(4\pi)^{d/2}}
y_c^{d/2}=\frac{4}{d(2\pi)^{d/2}}
 y_c^{d/4+1/2}\sum_{q=1}^{\infty}\frac{K_{d/2-1}(q\sqrt{y_c})}{q^{d/2-1}}.
\end{equation}
Using now that (see, e.g., \cite{GR73})
\begin{equation}\label{Kidentity}
K_{\nu+1}(z)=K_{\nu-1}(z)+\frac{2\nu}{z}K_{\nu}(z),
\end{equation}
from Eqs. (\ref{casamplsm}) and (\ref{criti}) we derive
\begin{equation}\label{adeltasm}
    \Delta=-\frac{2}{d(2\pi)^{d/2}}y_c^{d/4+1/2}\sum_{q=1}^{\infty}\frac{K_{d/2+1}(q\sqrt{y_c})}{q^{d/2-1}},
\end{equation}
with $\beta_cF_{{\rm Casimir}}(K_c,L_\perp)=(d-1)\ \Delta \
L_\perp^{-d}$. The exact value of $y_c$ and $\Delta$ in an
explicit form is only known for $d=3$. Then
\begin{equation}
\label{ayc} y=4\ln^2[(1+\sqrt{5})/2]
\end{equation}
(this value is well-known and seems that has been derived for the
first time in \cite{P72}), and \cite{D98}
\begin{equation}
\label{ay3} \Delta=-\frac{2\zeta(3)}{5\pi}.
\end{equation}
This is the only exactly know Casimir amplitude for a three
dimensional system. In \cite{D98} it has been shown (see there Eq.
(27)) that this value can also be written in the form
\begin{equation}
\label{d3L} \Delta=-\frac{1}{2\pi}\left[{\rm
Li}_3(e^{-\sqrt{y_c}})+\sqrt{y_c}\ {\rm
Li}_2(e^{-\sqrt{y_c}})+\frac{1}{6}y_c^{3/2} \right],
\end{equation}
where ${\rm Li}_p(z)=\sum_{k=1}^{\infty}z^k/k^p$ is the
polylogarithm function of order $p$. Taking into account that
$K(5/2,x)=\sqrt{\pi/2x}(1+2/x+3/x^2)\exp(-x)$ \cite{GR73} and Eq.
(\ref{yc}), one can easily check that the right hand side of Eq.
(\ref{deltasm}) can indeed be written in the form given in
(\ref{d3L}).

\subsubsection{Evaluation of the average value of the stress tensor}

Taking into account (\ref{mvt}), for the difference of the
finite-size and bulk internal energy densities $u=\langle {\cal
\hat{H}} \rangle$ and $u_b={\langle\cal \hat{H}}_b\rangle$, one
can easily derive from (\ref{freeenergy})
\begin{equation}\label{usm}
u-u_b=-\frac{1}{2}J(w-w_b)=-\frac{1}{4d}J(y-y_\infty)N^{-2}_\perp,
\end{equation}
where $y=2dwN_\perp^2$, and then from (\ref{cordef}), or
(\ref{freeenergy}), to obtain for the stress tensor (\ref{stsm})
that
\begin{eqnarray}\label{straverage}
\langle t_{\perp \perp}({\bf R}) \rangle &=&
\frac{d-1}{d}\frac{1}{N}\sum_{{\bf k}\in {\cal
B}_{\Lambda}}\frac{\cos(k_1 a_1)-\cos(k_d
a_d)}{d(1+w)-\sum_{j=1}^d \cos(k_j a_j)} \nonumber \\
&& -\frac{d-2}{4d} x_1(y-y_\infty) N_\perp^{-d},
\end{eqnarray}
where, we recall, $x_1=d^{-1}(K_c-K)N_\perp^{d-2}$ (see Eq.
(\ref{x1})). Using the identity (see Eq. (\ref{identitySP}))
\begin{equation}\label{identityI}
    \sum_{n=0}^{N-1}\cos\left(\frac{2\pi
    n}{N}\right) \exp\left[x \cos\left(\frac{2\pi
    n}{N}\right)\right]=N\sum_{q=-\infty}^\infty I_{q N}^{'}(x),
\end{equation}
where $I_\nu^{'}(x)=\frac{d}{dx}I_\nu(x)$, in the limit of a film
geometry, i.e. when $N_1, N_2, \cdots, N_{d-1} \rightarrow
\infty$, the above expression can be rewritten in the form
\begin{eqnarray}
\label{stavexact}
  \langle t_{\perp \perp}({\bf R}) \rangle &=&
\frac{2(d-1)}{d}\sum_{q=1}^{\infty}\int_0^\infty dx e^{-dwx}\left[e^{-x}I_0(x)\right]^{d-2} \nonumber \\
  &\times& e^{-2x}\left(I_0^{'}(x)I_{q
  N}(x)-I_0(x)I_{qN}^{'}(x)\right)\nonumber \\
 && -N_{\perp}^{-d} \frac{(d-2)}{4 d}x_1 (y-y_\infty).
\end{eqnarray}
It is worth to mention that till now no approximation in the
calculation of the of the average of the stress tensor operator
has been made. In order to obtain the scaling form of the above
expression such a step will be performed only now. Indeed, with
the help of the expansion \ref{SP85}
\begin{equation}\label{sp}
    I_\nu(x)=\frac{\exp(x-\nu^2/2x)}{\sqrt{2\pi x}} (1+\frac{1}{8 x}+\frac{9-32\nu^2}{2! (8x)^2}+ \ \cdots)
\end{equation}
one can set the above expression in scaling form
\begin{eqnarray}\label{stscalingform}
     \langle t_{\perp \perp}({\bf R}) \rangle &=& -
\frac{2(d-1)}{d(2\pi)^{d/2}}y^{(d+2)/4}\sum_{q=1}^{\infty}\frac{K_{1+d/2}(q\sqrt{y})}{q^{d/2-1}}N_\perp^{-d}\nonumber
\\
& & -N_{\perp}^{-d} \frac{(d-2)}{4 d}x_1 (y-y_\infty).
\end{eqnarray}
Now it only remains to show that the right-hand side of the above
equation is indeed equal to the right-hand side of Eq.
(\ref{cassm}) (for $b_\perp=b_\parallel$), which gives the Casimir
force calculated in a direct manner as a derivative of the
finite-size free energy with respect  of the size of the system.
In order to demonstrate that, let us first, with the help of
identity (\ref{Kidentity}), rewrite the above expression for the
stress tensor in the form
\begin{eqnarray}\label{astscalingformeq}
     \langle t_{\perp \perp}({\bf R}) \rangle &=& -\left\{
\frac{2(d-1)}{(2\pi)^{d/2}}\left[y^{d/4}\sum_{q=1}^{\infty}\frac{K_{d/2}(q\sqrt{y})}{q^{d/2}}\right.\right.
\nonumber
\\
& & \left.\left.
+\frac{1}{d}y^{(d+2)/4}\sum_{q=1}^{\infty}\frac{K_{d/2-1}(q\sqrt{y})}{q^{d/2-1}}
\right]\right.\nonumber \\
&& \left.+\frac{(d-2)}{4 d}x_1 (y-y_\infty)\right\}N_{\perp}^{-d}.
\end{eqnarray}
Next, from the bulk (\ref{sfsfeb}) and finite-size equations
(\ref{sfsfefinite}) of the spherical field  one directly derives
\begin{equation}\label{sfsfeby}
    -\frac{1}{4}x_1 y_\infty=\frac{\Gamma(1-d/2)}{2(4\pi)^{d/2}}
    y_\infty^{d/2},
\end{equation}
and
\begin{eqnarray}\label{sfsfefinitey}
   \lefteqn{ -\frac{1}{4}x_1 y=
   \frac{\Gamma(1-d/2)}{2(4\pi)^{d/2}}
    y^{d/2}}\\
    && + \frac{1}{(2\pi)^{d/2}}
y^{(d+2)/4}\sum_{q=1}^{\infty}\frac{K_{d/2-1}(q\sqrt{y})}{q^{d/2-1}},
\nonumber
\end{eqnarray}
wherefrom
\begin{eqnarray}
\label{help}
 \frac{1}{(2\pi)^{d/2}}
y^{(d+2)/4}\sum_{q=1}^{\infty}\frac{K_{d/2-1}(q\sqrt{y})}{q^{d/2-1}} = -\frac{1}{4} x_1 (y-y_\infty)\nonumber \\
 -\frac{\Gamma(1-d/2)}{2(4\pi)^{d/2}}
 \left(y^{d/2}-y_\infty^{d/2}\right).\ \ \ \ \ \ \ \ \
\end{eqnarray}
Inserting now (\ref{help}) in (\ref{stscalingformeq}) we conclude,
that, indeed,
\begin{equation}\label{asmfinal}
\langle t_{\perp \perp}({\bf R}) \rangle=\beta F_{{\rm Casimir}}
\end{equation}
for the spherical model.

\subsubsection{Evaluation of the variance of the stress tensor}
\label{vstsmsection}

If $\Delta \zeta$ is the variance of the random variable $\zeta$,
i.e. $\Delta \zeta=<(\zeta-<\zeta>)^2>=<\zeta^2>-<\zeta>^2$, then
at $T=T_c$
\begin{eqnarray}\label{dtany}
\Delta \sum_{{\bf R}\in \Lambda}\tilde{t}_{\perp,\perp}({\bf
R})&=& \frac{\partial^2}{\partial \lambda^2}\left[\ln \sum
e^{-\beta {\cal H}(\lambda)} \right]
\\
\nonumber  &=& N_\perp N_\parallel^{d-1}\frac{\partial^2}{\partial
\lambda^2}[-\beta \tilde{f}(T_c,\lambda)]
\end{eqnarray}
(for the definition of ${\cal H}(\lambda)$) and
$\tilde{f}(T,\lambda)$ see Eqs. (\ref{htt})-(\ref{tt}). Comparing
now Eqs. (\ref{tt}) and (\ref{stsm}), we conclude that, at
$T=T_c$, in the case of a spherical model
\begin{equation}\label{vsm}
\Delta \sum_{{\bf R}\in \Lambda}t_{\perp,\perp}({\bf R}) =
\frac{4}{d^2} N_\perp N_\parallel^{d-1}\frac{\partial^2}{\partial
\lambda^2}[-\beta \tilde{f}(T_c,\lambda)].
\end{equation}
The finite-size free energy density of the anisotropic system is
given in Eq. (\ref{finitesizefreeenergy}), where the anisotropy is
characterized by the constants
\begin{equation}
\label{bpar}
b_1=b_2=\cdots=b_{d-1}=b_\parallel=\frac{J_\parallel}{(d-1)J_\parallel+J_\perp}=\frac{1+\lambda}{d},
\end{equation}
and
\begin{equation}\label{bperp}
b_d=b_\perp=\frac{J_\perp}{(d-1)J_\parallel+J_\perp}=\frac{1}{d}-\frac{d-1}{d}\lambda.
\end{equation}
Let us now first note that
\begin{equation}\label{K}
K=\beta \hat{J}({\bf
0})=2\beta\left[(d-1)J_\parallel+J\perp\right]=2\beta Jd
\end{equation}
does not depend on $\lambda$ and that
\begin{equation}\label{db}
\frac{\partial b_\parallel}{\partial \lambda}=\frac{1}{d}, \qquad
\frac{\partial b_\perp}{\partial \lambda}=-\frac{d-1}{d}.
\end{equation}
It is clear from Eqs. (\ref{finitesizefreeenergy}) and
(\ref{excess})) that the contributions to the variance of the
stress tensor stemming from the "finite size"  and the singular
"bulk" parts will be of the order of $(N_\perp
N_\parallel^{d-1})/N_\perp^{d}$, wile that one from the "bulk"
regular part will be of the order of $N_\perp N_\parallel ^{d-1}$.
Because of that the leading contributions will be nonuniversal. In
addition to them one will have also universal corrections, but we
will neglect them in the current treatment and will deal only with
the leading-order behavior of the variance of the Casimir force.
Then, from  Eq. (\ref{finitesizefreeenergy}), one has
\begin{equation}\label{vstsm}
\Delta \sum_{{\bf R}\in \Lambda}t_{\perp,\perp}({\bf R}) \sim
\frac{4}{d^2} N_\perp N_\parallel^{d-1}\frac{\partial^2}{\partial
\lambda^2}[-U_b(0|{\bf b}],
\end{equation}
where $U_b(0|{\bf b})\equiv U_b(0)$ is defined in Eq. (\ref{ub0}).
Having in mind Eq. (\ref{db}), it is easy to show that
\begin{equation}\label{dUb0}
\frac{\partial}{\partial \lambda}U_b(0|{\bf
b})=\frac{d-1}{d}\left(\frac{\partial}{\partial b_\parallel}
U_b(0|{\bf b}) -\frac{\partial}{\partial b_\perp} U_b(0|{\bf b})
\right),
\end{equation}
wherefrom one immediately derives
\begin{equation}\label{sdub0}
\frac{\partial^2}{\partial \lambda^2}U_b(0|{\bf
b})=\frac{d-1}{d}\left(\frac{\partial^2}{\partial b_\parallel^2}
U_b(0|{\bf b}) -\frac{\partial^2}{\partial b_\perp\partial
b_\parallel} U_b(0|{\bf b})\right).
\end{equation}
Performing now the calculations, from (\ref{ub0}) and
(\ref{sdub0}), we obtain
\begin{eqnarray}\label{sdub0final}
\frac{\partial^2}{\partial \lambda^2}U_b(0|{\bf
b})&=&\frac{1}{2}d(d-1)\int_0^\infty dx x e^{-d x}I_0^{d-2}(x) \\
\nonumber && \times
\left\{I_1^2(x)-\frac{1}{2}I_0(x)(I_0(x)+I_2(x))\right\},
\end{eqnarray}
which leads to the following result
\begin{eqnarray}\label{vstsmfinal} \nonumber
 \Delta \sum_{{\bf R}\in \Lambda}t_{\perp,\perp}({\bf R}) &\simeq&
-\frac{2(d-1)}{d} N_\perp N_\parallel^{d-1}\int_0^\infty dx  x I_0^{d-2}(x) \\
 \times
\left\{I_1^2(x)\right. &-& \left.
\frac{1}{2}I_0(x)(I_0(x)+I_2(x))\right\} e^{-d x},
\end{eqnarray}
for the variance of the Casimir force in the spherical model at
$T=T_c$. This will be also the leading result everywhere in the
critical region.
A numerical evaluation of Eq. (\ref{vstsmfinal})
gives
\begin{equation}\label{avsmnumerical}
\Delta t_ {\perp,\perp} \equiv \Delta \sum_{{\bf R}\in
\Lambda}t_{\perp,\perp}({\bf R}) \simeq 0.107\  N_\perp
N_\parallel^2.
\end{equation}


\end{document}